\documentclass[Crown,times]{sagej}

\usepackage{moreverb,url}

\usepackage[colorlinks,bookmarksopen,bookmarksnumbered,citecolor=red,urlcolor=red]{hyperref}

\def\volumeyear{2023}
\graphicspath{{./LCE-chargepump-figs/}}
\begin{document}

\runninghead{L. A. Mihai}

\title{A theoretical model for power generation via liquid crystal elastomers}

\author{L. Angela Mihai\affilnum{1}}

\affiliation{\affilnum{1}School of Mathematics, Cardiff University,  Cardiff, UK}

\corrauth{L. Angela Mihai, School of Mathematics, Cardiff University,  Cardiff, UK.}

\email{MihaiLA@cardiff.ac.uk}

\begin{abstract}
Motivated by the need for new materials and green energy production and conversion processes, a class of mathematical models for liquid crystal elastomers integrated within a theoretical charge pump electrical circuit is considered. The charge pump harnesses the chemical and mechanical properties of liquid crystal elastomers transitioning from the nematic to isotropic phase when illuminated or heated to generate higher voltage from a lower voltage supplied by a battery. For the material constitutive model, purely elastic and neoclassical-type strain energy densities applicable to a wide range of monodomain nematic elastomers are combined, while elastic and photo-thermal responses are decoupled to make the investigation analytically tractable. By varying the model parameters of the elastic and neoclassical terms, it is found that liquid crystal elastomers are more effective than rubber when used as dielectric material within a charge pump capacitor. 

\end{abstract}


\maketitle

%%%%%%%%%%%%%%%%%%%%%%%%%%%%%%%%%%%%%%%%%%%%%%%%%%%%%%%%%%%%
%%%%%%%%%%%%%%%%%%%%  NEW SECTION   %%%%%%%%%%%%%%%%%%%%%%%%
%%%%%%%%%%%%%%%%%%%%%%%%%%%%%%%%%%%%%%%%%%%%%%%%%%%%%%%%%%%%
\section{Introduction}\label{NLC:sec:intro:cp}

To end the use of fossil fuels, more materials that enable green energy production and conversion processes are sought and developed \cite{Gallo:2016:etal,Renewables:2022}. In particular, flexible energy harvesters made of rubber-like materials demonstrate great potential for generating low carbon renewable energy in emerging technologies \cite{Lu:2020:etal,Moretti:2020:etal}. 

This paper considers a liquid crystal elastomer (LCE) incorporated in a theoretical charge pump electrical circuit. Charge pumps are convenient and economical devices that use capacitors to generate higher voltages from a lower voltage supplied by a source battery. The simplest capacitor consists of two parallel plate electrical conductors separated by air or an insulating material known as the dielectric. The plates are connected to two terminals, which can be wired into an electric circuit. When the performance of a capacitor changes by altering the distance between plates or the amount of plate surface area, a variable capacitor is achieved. 

%%%%%%%%%%%%%%%
\begin{figure}[htbp] 
	\begin{center}
		\includegraphics[width=0.5\textwidth]{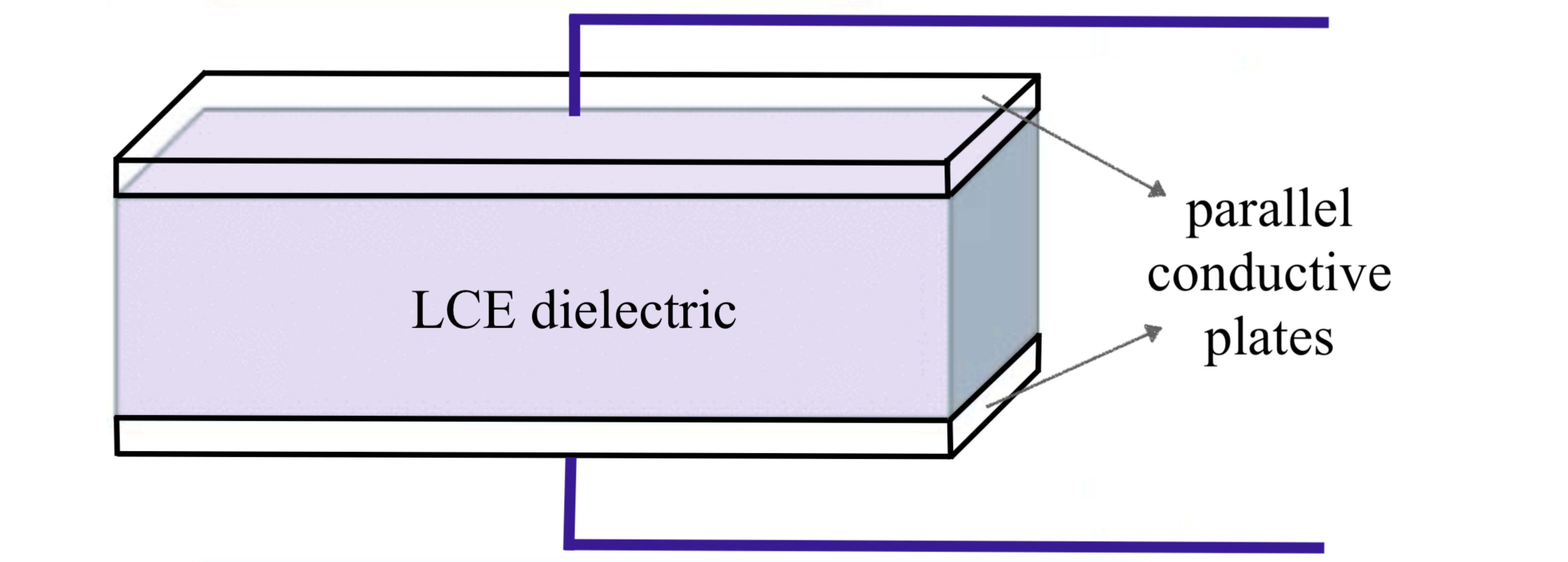}
		\caption{Schematic of parallel plate capacitor with LCE dielectric between two compliant electrodes.}\label{NLC:fig:capacitor}
	\end{center}
\end{figure}
%%%%%%%%%%%%%%

%%%%%%%%%%%%%%%
\begin{figure}[htbp] 
	\begin{center}
		\includegraphics[width=0.99\textwidth]{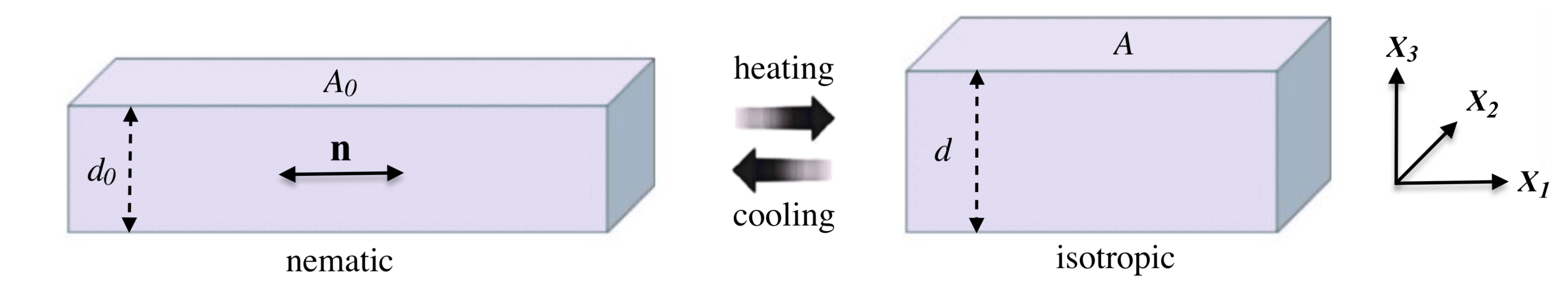}
		\caption{Reversible natural deformation of LCE dielectric under nematic-isotropic phase transition caused by temperature variation. The thickness $d=\lambda d_{0}$ and surface area $A$ of the LCE change, while the volume is preserved, i.e., $\Omega=A_{0}d_{0}=Ad$. In the nematic phase, the director $\textbf{n}$ for liquid crystal orientation is aligned in the first Cartesian direction, parallel to the surface, while in the isotropic phase, liquid crystal molecules are randomly oriented.}\label{NLC:fig:nemiso}
	\end{center}
\end{figure}
%%%%%%%%%%%%%%%

LCEs are top candidates for dielectric material because they are capable of large strain deformations which are reversible and repeatable under natural stimuli like heat and light \cite{deJeu:2012,Warner:2007:WT}. This is due to their unique molecular architecture combining the flexibility of polymeric networks with liquid crystal self-organisation \cite{deGennes:1993:dGP}. In Figure~\ref{NLC:fig:capacitor}, a capacitor with LCE dielectric between two compliant electrodes is presented schematically. Figure~\ref{NLC:fig:nemiso} depicts the LCE nematic-isotropic phase transition under thermal stimuli. Light-induced shape changes for LCEs containing photoisomerizing dye molecules can be represented in a similar manner.

A hypothetical charge pump which converts solar heat into DC electricity was proposed in \cite{Hiscock:2011:HWPM}. In that study, the LCE was described by the neoclassical model  \cite{Bladon:1994:BTW,Warner:1988:WGV,Warner:1991:WW} and elastic and thermal responses were decoupled to make the theoretical model analytically tractable.
	
In this paper, purely elastic and neoclassical-type strain-energy densities are combined. The resulting composite model is applicable to a wide range of nematic elastomers and can be reduced to either the neo-Hookean model for rubber \cite{Treloar:1944} or the neoclassical model for ideal LCEs.  As in \cite{Hiscock:2011:HWPM}, the elastic deformation and photo-thermal responses are decoupled. Then, if heat or light is absorbed, the equilibrium uniaxial order parameter can be determined by minimising the Landau-de Gennes approximation of the nematic energy density \cite{deGennes:1993:dGP} or a Maier-Saupe mean field model function \cite{Bai:2020:BB,Corbett:2006:CW,Corbett:2008:CW,Maier:1959:MS}, respectively. By varying the model parameters of the elastic and neoclassical terms, it is found that LCEs can be more effective than rubber when used as dielectric material within a charge pump capacitor. Moreover, if the LCE is pre-stretched perpendicular to the director and instabilities such as shear striping or wrinkling are avoided, then the capacitor becomes more efficient in raising the voltage supplied by the source battery.

%%%%%%%%%%%%%%%%%%%%%%%%%%%%%%%%%%%%%%%%%%%%%%%%%%%%%%%%%%%%
%%%%%%%%%%%%%%%%%%%%  NEW SECTION   %%%%%%%%%%%%%%%%%%%%%%%%
%%%%%%%%%%%%%%%%%%%%%%%%%%%%%%%%%%%%%%%%%%%%%%%%%%%%%%%%%%%%
\section{The charge pump circuit}\label{NLC:sec:circuit:cp}

The electrical energy potential stored by a capacitor, known as capacitance, is equal to $C=q/V$, where $q$ is the magnitude of the charge stored when the voltage across the capacitor is equal to $V$, and is measured in farads (F). Capacitance depends on both the geometry and the materials that the capacitor is made of. 

%%%%%%%%%%%%%%%
\begin{figure}[htbp] 
	\begin{center}
		\includegraphics[width=0.3\textwidth]{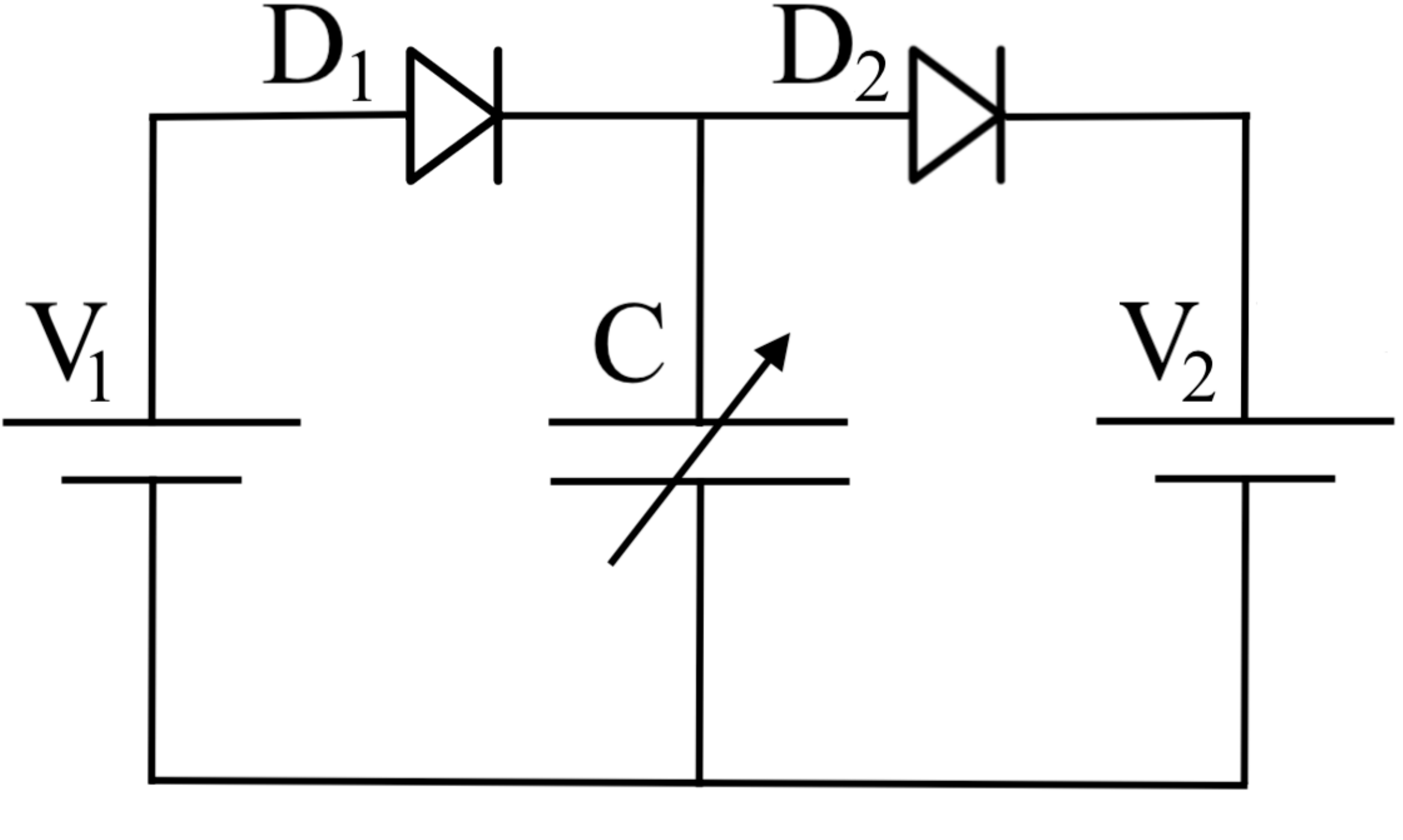}
		\caption{Charge pump electrical circuit with a single-cell supply battery of lower voltage $V_{1}$ and a variable capacitor with capacitance $C$ that attains a higher voltage $V_{2}$ \cite{Hiscock:2011:HWPM}. }\label{NLC:fig:circuit1}
	\end{center}
%\end{figure}
%%%%%%%%%%%%%%%
%%%%%%%%%%%%%%%
%\begin{figure}[htbp] 
	\begin{center}
		\includegraphics[width=0.3\textwidth]{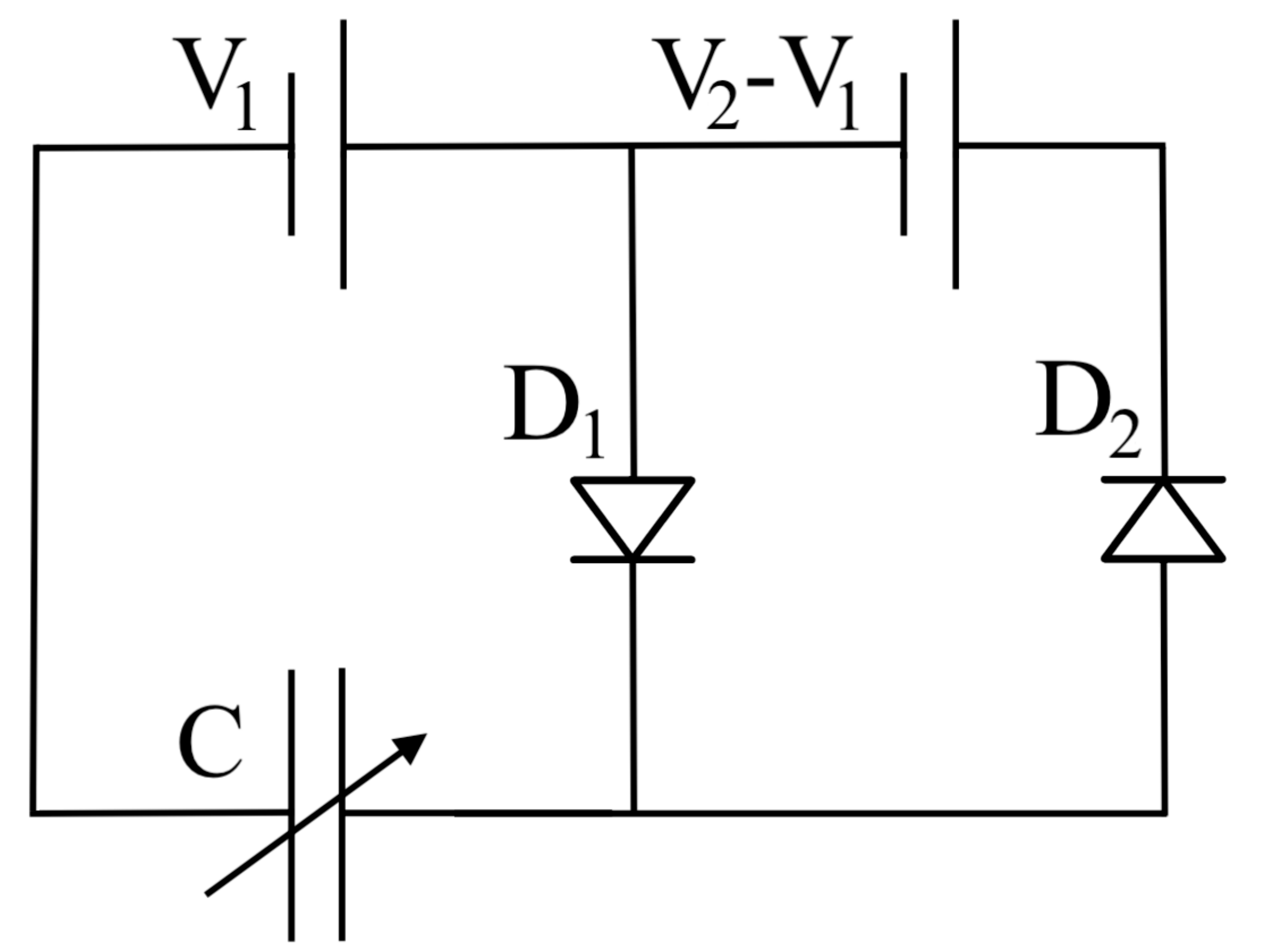}
		\caption{Charge pump electrical circuit with a single-cell supply battery of lower voltage $V_{1}$ and a variable capacitor with capacitance $C$ that attains a higher voltage $V_{2}$ and recharges the supply battery \cite{Hiscock:2011:HWPM}. }\label{NLC:fig:circuit2}
	\end{center}
\end{figure}
%%%%%%%%%%%%%%%
%%%%%%%%%%%%%%%
\begin{figure}[htbp] 
	\begin{center}
		\includegraphics[width=0.45\textwidth]{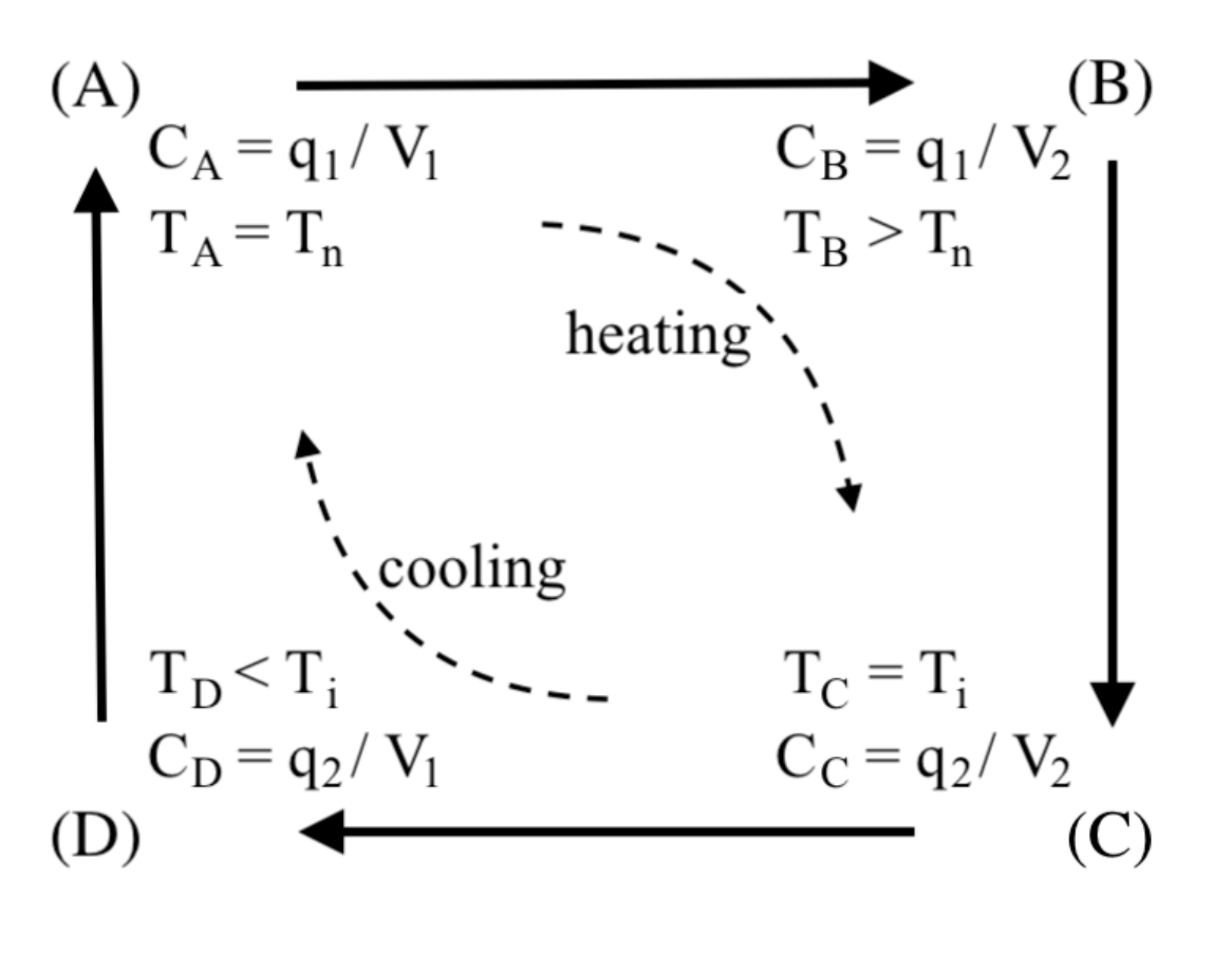}
		\caption{Operating cycle of variable capacitor with the dielectric made of LCE generating a higher voltage $V_{2}$ from a lower voltage $V_{1}$ \cite{Hiscock:2011:HWPM}.}\label{NLC:fig:cpump}
	\end{center}
\end{figure}
%%%%%%%%%%%%%%%

Figure~\ref{NLC:fig:circuit1} represents schematically a charge pump electrical circuit where the voltage $V_{1}$ supplied by an external source (battery) is raised to voltage $V_{2}$ by using a variable capacitor with capacitance $C$. The diodes prevent backflow of charge and act as voltage activated switches. The alternative circuit illustrated in Figure~\ref{NLC:fig:circuit2} allows for the supply battery to be recharged. Figure~\ref{NLC:fig:cpump} shows the operating cycle of a variable capacitor with LCE dielectric containing the following states \cite{Hiscock:2011:HWPM}:
\begin{itemize}
	\item[(A)] The LCE is at the lowest temperature corresponding to the nematic state, $T_{n}=T_{A}$, the input voltage from a supply battery is $V_{1}$ and the capacitor is charged to the initial charge $q_{1}$. At this state, the capacitance is equal to $C_{n}=C_{A}=q_{1}/V_{1}$;
	\item[(B)] The temperature rises to $T_{B}>T_{A}$, so the capacitance decreases to $C_{B}<C_{A}$, while the charge remains equal to $q_{1}$. Thus the voltage across the capacitor increases to $V_{2}=q_{1}/C_{B}>q_{1}/C_{A}=V_{1}$;
	\item[(C)] The temperature continues to increase to the isotropic state, $T_{i}=T_{C}>T_{B}$, hence the capacitance further decrease to $C_{i}=C_{C}<C_{B}$, but the voltage remains equal to $V_{2}$, so the charge decreases to $q_{2}=C_{C}V_{2}<C_{B}V_{2}=q_{1}$;
	\item[(D)] The temperature drops to $T_{D}<T_{i}$, while the charge remains equal to $q_{2}$ and the capacitance increases to $C_{D}=q_{2}/V_{1}>q_{2}/V_{2}=C_{C}$. Upon further cooling to the initial temperature $T_{n}$, the battery with voltage $V_{1} $ charges the capacitor to the initial charge state $q_{1}$ and the cycle can be repeated.
\end{itemize}
In the above notation and throughout the paper, indices `$n$' and `$i$' indicate a nematic or isotropic phase, respectively.
 
During one cycle, the external source provides an electrical energy $W_{in}=V_{1}\left(q_{1}-q_{2}\right)$ and produces $W_{out}=V_{2}\left(q_{1}-q_{2}\right)$. Thus the net output generated by this cycle is equal to
\begin{equation}
W=W_{out}-W_{in}=\left(V_{2}-V_{1}\right)\left(q_{1}-q_{2}\right)=-C_{n}V_{1}^2\left(\frac{C_{i}}{C_{B}}-1\right)\left(\frac{C_{n}}{C_{B}}-1\right).
\end{equation}
By defining the capacitance ratio
\begin{equation}\label{NLC:eq:xi:CnCi}
\xi=\frac{C_{n}}{C_{i}}>1,
\end{equation}
the maximum generated output per cycle is equal to
\begin{equation}\label{NLC:eq:Wm:xi}
W_{m}(\xi)=C_{n}V_{1}^2\frac{(\xi-1)^2}{4\xi}.
\end{equation}
This is attained when
\begin{equation}
C_{B}=\frac{2C_{n}C_{i}}{C_{n}+C_{i}},
\end{equation}
or equivalently, when
\begin{equation}\label{NLC:eq:xi:V1V2}
\frac{V_{2}}{V_{1}}=\frac{\xi+1}{2}.
\end{equation}

For the LCE transitioning from a nematic to an isotropic phase and vice versa, changes in light instead of temperature can be used as well.

%%%%%%%%%%%%%%%%%%%%%%%%%%%%%%%%%%%%%%%%%%%%%%%%%%%%%%%%%%%%
%%%%%%%%%%%%%%%%%%%%  NEW SECTION   %%%%%%%%%%%%%%%%%%%%%%%%
%%%%%%%%%%%%%%%%%%%%%%%%%%%%%%%%%%%%%%%%%%%%%%%%%%%%%%%%%%%%
\section{The LCE strain-energy function}\label{NLC:sec:model:cp}

To describe the incompressible nematic LCE, the following form of the elastic strain-energy density function is assumed \cite{Conti:2002a:CdSD,Fried:2004:FS,Fried:2005:FS,Fried:2006:FS,Lee:2021:LB,Mihai:2020a:MG,Mihai:2021a:MG},
\begin{equation}\label{NLC:eq:Wel:cp}
\mathcal{W}^{(el)}(\textbf{F},\textbf{n})=\mathcal{W}^{(1)}\left(\textbf{F}\right)+\mathcal{W}^{(2)}\left(\textbf{G}^{-1}\textbf{F}\textbf{G}_{0}\right),
\end{equation}
where $\textbf{F}$ denotes the deformation gradient from the reference cross-linking state, such that $\det\textbf{F}=1$, while $\textbf{n}$ is a unit vector for the localised direction of uniaxial nematic alignment in the present configuration and is termed ``the director''.  The first term on the right-hand side of equation \eqref{NLC:eq:Wel:cp} represents the strain-energy density associated with the overall macroscopic deformation, and the second term is the strain-energy density of the polymer microstructure. In the second term, $\textbf{G}_{0}$ and $\textbf{G}$ denote the natural (or spontaneous) deformation tensor in the reference and current configuration, respectively \cite{Finkelmann:2001:FGW,Warner:2007:WT}.

Assuming the LCE to be intrinsically uniaxial, the natural deformation tensor takes the form
\begin{equation}\label{NLC:eq:G}
\textbf{G}=a^{-1/6}\textbf{I}+\left(a^{1/3}-a^{-1/6}\right)\textbf{n}\otimes\textbf{n},
\end{equation}
where $\textbf{I}=\mathrm{diag}(1,1,1)$ is the identity tensor and 
\begin{equation}\label{NLC:eq:aQ}
a=\frac{1+2Q}{1-Q}
\end{equation}
denotes the natural shape parameter, with $Q$ representing the scalar uniaxial order of the liquid crystal mesogens ($Q=1$ corresponds to perfect nematic order, while $Q=0$ is for the case when the mesogens are randomly oriented). In the reference configuration, $\textbf{G}$ is replaced by $\textbf{G}_{0}$, with $\textbf{n}_{0}$, $a_{0}$ and $Q_{0}$ instead of $\textbf{n}$, $a$ and  $Q$, respectively. Setting $\textbf{n}_{0}=\textbf{n}=[1,0,0]^\mathrm{T}$, the natural deformation tensors take the form
\begin{equation}
\textbf{G}_{0}=\mathrm{diag}\left(a_{0}^{1/3},a_{0}^{-1/6},a_{0}^{-1/6}\right),\qquad
\textbf{G}=\mathrm{diag}\left(a^{1/3},a^{-1/6},a^{-1/6}\right).
\end{equation}

The components of phenomenological model given by equation \eqref{NLC:eq:Wel:cp} are defined as follows,
\begin{equation}\label{NLC:eq:W:nh1}
\mathcal{W}^{(1)}(\lambda_{1},\lambda_{2},\lambda_{3})=\frac{\mu^{(1)}}{2}\left(\lambda_{1}^{2}+\lambda_{2}^{2}+\lambda_{3}^{2}\right),
\end{equation}
where $\mu^{(1)}>0$ is a constant independent of the deformation and $\{\lambda_{1}^2,\lambda_{2}^2,\lambda_{3}^2\}$ are the eigenvalues of the tensor $\textbf{F}^{T}\textbf{F}$, such that $\lambda_{1}\lambda_{2}\lambda_{3}=1$, and
\begin{equation}\label{NLC:eq:W:nh2}
\mathcal{W}^{(2)}(\alpha_{1},\alpha_{2},\alpha_{3})=\frac{\mu^{(2)}}{2}\left(\alpha_{1}^{2}+\alpha_{2}^{2}+\alpha_{3}^{2}\right),
\end{equation}
where $\mu^{(2)}>0$ is a constant independent of the deformation and $\{\alpha_{1}^2,\alpha_{2}^2,\alpha_{3}^2\}$ are the eigenvalues of the elastic Cauchy-Green tensor $\textbf{A}^{T}\textbf{A}$, with the local elastic deformation tensor $\textbf{A}=\textbf{G}^{-1}\textbf{F}\textbf{G}_{0}$, such that $\alpha_{1}\alpha_{2}\alpha_{3}=1$. Note that these components are derived from the classical neo-Hookean model for rubber \cite{Treloar:1944}.

The composite model defined by equation \eqref{NLC:eq:Wel:cp} thus takes the form
\begin{equation}\label{NLC:eq:Wel:cp:nh}
\mathcal{W}^{(el)}=\frac{\mu^{(1)}}{2}\left(\lambda_{1}^{2}+\lambda_{2}^{2}+\lambda_{3}^{2}\right)
+\frac{\mu^{(2)}}{2}\left(\alpha_{1}^{2}+\alpha_{2}^{2}+\alpha_{3}^{2}\right),
\end{equation}
and has the shear modulus in the infinitesimal strain equal to $\mu=\mu^{(1)}+\mu^{(2)}>0$. This strain-energy function reduces to the neoclassical model for LCEs when $\mu^{(1)}=0$ \cite{Bladon:1994:BTW,Warner:1988:WGV,Warner:1991:WW} and to the neo-Hookean model for rubber when $\mu^{(2)}=0$ \cite{Treloar:1944}.

%%%%%%%%%%%%%%%%%%%%%%%%%%%%%%%%%%%%%%%%%%%%%%%%%%%%%%%%%%%%
\subsection{Photo-thermal responses}\label{NLC:sec:model:pr}

If azobenzene mesogens are embedded in the nematic elastomeric network, then, when photons are absorbed, the so-called \textit{Weigert effect} \cite{Ebralidze:1995,Kakichashvili:1974} occurs where the dye molecules change from straight \emph{trans}- to bent \emph{cis}-isomers, causing a reduction in the nematic order. To account for photo-thermal deformation of a dielectric LCE, we adopt the following modified Maier-Saupe mean field model \cite{Bai:2020:BB,Corbett:2006:CW,Corbett:2008:CW,Maier:1959:MS},
\begin{equation}\label{NLC:eq:Wms:cp}
\mathcal{W}^{(ms)}(Q,\textbf{n})=\mu^{(ms)}\left(1-c(Q,\textbf{n})\right)\left[g^{-1}(Q)Q-\log Z(Q)-\frac{\tilde{J}}{2}Q^2\left(1-c(Q,\textbf{n})\right)\right],
\end{equation}
where $Q$ and $\textbf{n}$ are defined as before, $\mu^{(ms)}=N^{(ms)}kT$, with $N^{(ms)}$ the total number of mesogens per unit volume and $kT$ the temperature per unit of energy, $\tilde{J}$ represents the average interaction between two mesogens in the unit of energy, and $c(\cdot,\cdot)$ denotes the fractional number of \emph{cis} molecules. 
\begin{itemize}
	\item If the light is \emph{polarised} and $\alpha$ denotes the angle between nematic director $\textbf{n}$ and the light polarisation, then
	\begin{equation}\label{NLC:eq:c:pol}
	c(Q,\textbf{n})=f\frac{I\left[1+Q\left(3\cos^2\alpha-1\right)\right]}{3+I\left[1+Q\left(3\cos^2\alpha-1\right)\right]},
	\end{equation}
	where $f$ is the fraction of photo-active mesogens and $I$ is the non-dimensional homogeneous light intensity.
	
	\item If the light is \emph{unpolarised} and $\alpha$ is the angle between nematic director $\textbf{n}$ and the light beam direction, then
	\begin{equation}\label{NLC:eq:c:unpol}
	c(Q,\textbf{n})=f\frac{I\left[1-(Q/2)\left(3\cos^2\alpha-1\right)\right]}{3+I\left[1-(Q/2)\left(3\cos^2\alpha-1\right)\right]}.
	\end{equation}
\end{itemize}
The expressions for the functions $g(\cdot)$ and $Z(\cdot)$ are, respectively,
\begin{equation}
g(x)=-\frac{1}{2}-\frac{1}{2x}+\frac{1}{2x}\sqrt{\frac{3x}{2}}\frac{\exp\left(3x/2\right)}{\int_{0}^{\sqrt{\left(3x/2\right)}}\exp\left(y^2\right)\text{d}y}
\end{equation}
and
\begin{equation}
Z(Q)=\frac{\exp\left(g^{-1}(Q)\right)}{1+g^{-1}(Q)\left(1+2Q\right)}.
\end{equation}
In the absence of light, $I=0$ and the energy function defined by \eqref{NLC:eq:Wms:cp} reduces to
\begin{equation}\label{NLC:eq:Wms:0}
\mathcal{W}^{(ms)}(Q,\textbf{n})=\mu^{(ms)}\left(g^{-1}(Q)Q-\log Z(Q)-\frac{\tilde{J}}{2}Q^2\right).
\end{equation}
The ratio between photoactive and non-photoactive mesogens can also be taken into account \cite{Corbett:2006:CW,Corbett:2008:CW,Goriely:2022:GMM}.

%%%%%%%%%%%%%%%%%%%%%%%%%%%%%%%%%%%%%%%%%%%%%%%%%%%%%%%%%%%%
%%%%%%%%%%%%%%%%%%%%  NEW SECTION   %%%%%%%%%%%%%%%%%%%%%%%%
%%%%%%%%%%%%%%%%%%%%%%%%%%%%%%%%%%%%%%%%%%%%%%%%%%%%%%%%%%%%
\section{Energy conversion}\label{NLC:sec:capacitor:cp}

At state (A), the LCE dielectric is considered either in its natural configuration or pre-stretched perpendicular or parallel to the nematic director, so that the surface area is increased and the distance between plates is reduced, hence the initial capacitance increases. In all cases, the LCE can be actuated by illumination or heating.

Assuming  $\mu\ll\mu^{(ms)}$, the energy functions $\mathcal{W}^{(el)}$ and $\mathcal{W}^{(ms)}$ described by equations \eqref{NLC:eq:Wel:cp:nh} and \eqref{NLC:eq:Wms:cp}, respectively, can be treated separately, and the equilibrium scalar order parameter $Q$ obtained by minimising the function $\mathcal{W}^{(ms)}$. Experimental results on photo-thermal shape changes in LCEs are reported in \cite{Finkelmann:2001:FNPW,Guo:2022:etal,Yu:2003:YN}. In \cite{Goriely:2022:GMM}, a general theoretical model for photo-mechanical responses in nematic-elastic rods is presented. Photoactive LCE beams under illumination are modelled in \cite{Norouzikudiani:2023:NLDeS}. Reviews of various light-induced mechanical effects can be found in \cite{Ambulo:2020:etal,McCracken:2020:etal,Warner:2020,Wen:2020:etal}. 

Similarly, when heat instead of light is absorbed \cite{Hiscock:2011:HWPM}, the uniaxial order parameter $Q$ can be determined by minimising the following Landau-de Gennes approximation of the nematic energy density
\begin{equation}\label{NLC:eq:W:lc}
\mathcal{W}^{(lc)}(Q)=\frac{\texttt{A}}{2}Q^2-\frac{\texttt{B}}{3}Q^3+\frac{\texttt{C}}{4}Q^4,
\end{equation}
where $\texttt{A}, \texttt{B}, \texttt{C}$ are material constants, with $\texttt{A}=\texttt{A}(T)$ depending on temperature \cite[p.~15]{Warner:2007:WT}. For nematic LCEs, the contribution of the above function to the total strain-energy density including both the isotropic elastic energy and the nematic energy functions was originally analysed in \cite{Finkelmann:2001:FGW} and more recently in \cite{Mihai:2021:MWGG}. 

%%%%%%%%%%%%%%%%%%%%%%%%%%%%%%%%%%%%
\subsection{Natural deformation}

When the LCE is in its natural configuration at state (A), as the capacitor is connected to the source battery, the total energy function of the system takes the following form (see also \cite{Hiscock:2011:HWPM}),
\begin{equation}\label{NLC:eq:W:lambda1lambda}
\mathcal{W}(\lambda_{1},\lambda)=\frac{\mu^{(1)}}{2}\left(\lambda_{1}^{2}+\lambda_{1}^{-2}\lambda^{-2}+\lambda^{2}\right)+\frac{\mu^{(2)}}{2}a^{1/3}a_{0}^{-1/3}\left(\lambda_{1}^{2}a^{-1}a_{0}+\lambda_{1}^{-2}\lambda^{-2}+\lambda^{2}\right)
-\frac{C V^2}{2\Omega},
\end{equation}
where the last term represents the electrical energy per unit volume. Here, $\Omega=Ad$ is the volume of the dielectric, with $A$ the surface area and $d=\lambda d_{0}$ the distance between the conductive plates, $V$ is the voltage across the capacitor, and $C$ is the capacitance given by
\begin{equation}\label{NLC:eq:C}
C=\frac{\varepsilon_{\perp}\varepsilon_{0\perp}A}{d}=\frac{\varepsilon_{\perp}\varepsilon_{0\perp}\Omega}{d_{0}^2\lambda^2},
\end{equation}
where $\varepsilon_{0\perp}$ is the permittivity for the perfectly nematic phase and $\varepsilon_{\perp}$ is the relative permittivity when the director is perpendicular to the electric field. The ``$\perp$'' notation stands for the electric field being applied perpendicular to the nematic director.

We denote by $V_{m}$ the voltage where the total energy is comparable to the stored elastic energy, such that
\begin{equation}\label{NLC:eq:Vm}
V_{m}^2=\frac{\mu d_{0}^2}{\varepsilon_{0\perp}}.
\end{equation}
Minimising the total energy function described by equation \eqref{NLC:eq:W:lambda1lambda} with respect to $\lambda_{1}$ gives
\begin{equation}\label{NLC:eq:W:lambda}
\begin{split}
&\mathcal{W}(\lambda)
=\frac{\mu^{(1)}}{2}\left[\left(\frac{\mu^{(1)}/\mu^{(2)}+a^{1/3}/a_{0}^{1/3}}{\mu^{(1)}/\mu^{(2)}+a_{0}^{2/3}/a^{2/3}}\right)^{1/2}\lambda^{-1}+\left(\frac{\mu^{(1)}/\mu^{(2)}+a^{1/3}/a_{0}^{1/3}}{\mu^{(1)}/\mu^{(2)}+a_{0}^{2/3}/a^{2/3}}\right)^{-1/2}\lambda^{-1}+\lambda^{2}\right]\\
&+\frac{\mu^{(2)}}{2}\frac{a^{1/3}}{a_{0}^{1/3}}\left[\left(\frac{\mu^{(1)}/\mu^{(2)}+a^{1/3}/a_{0}^{1/3}}{\mu^{(1)}/\mu^{(2)}+a_{0}^{2/3}/a^{2/3}}\right)^{1/2}\lambda^{-1}a^{-1}a_{0}+\left(\frac{\mu^{(1)}/\mu^{(2)}+a^{1/3}/a_{0}^{1/3}}{\mu^{(1)}/\mu^{(2)}+a_{0}^{2/3}/a^{2/3}}\right)^{-1/2}\lambda^{-1}+\lambda^{2}\right]\\
&-\frac{\mu}{2}\varepsilon_{\perp}V^2V_{m}^{-2}\lambda^{-2}.
\end{split}
\end{equation}

At the initial state (A), where the LCE exhibits nematic alignment, there is no light, i.e., $I=0$, and minimising the energy function defined by equation \eqref{NLC:eq:Wms:0} with respect to $Q$ yields the optimal value $Q_{0}$. At this state also, $V=V_{1}$ and $a=a_{0}$, hence the function given by equation \eqref{NLC:eq:W:lambda} becomes
\begin{equation}\label{NLC:eq:Wn}
\mathcal{W}_{n}(\lambda)=\frac{\mu}{2}\left(\lambda^{2}+2\lambda^{-1}\right)-\frac{\mu}{2}v\lambda^{-2}, 
\end{equation}
where
\begin{equation}\label{NLC:eq:v}
v=\frac{\varepsilon_{1}V_{1}^2}{V_{m}^2}
\end{equation}
denotes the operating voltage. By solving for $\lambda=\lambda_{n}$ and $v=v_{n}$ the following system of equations,
\begin{equation}
\frac{\partial \mathcal{W}_{n}}{\partial\lambda}=0,\qquad \frac{\partial^{2} \mathcal{W}_{n}}{\partial\lambda^{2}}=0,
\end{equation}
we obtain
\begin{equation}\label{NLC:eq:lambdan:vn}
\lambda_{n}=\frac{1}{4^{1/3}}\qquad\mbox{and}\qquad v_{n}=\frac{3\lambda_{n}}{4}=\frac{3}{4^{4/3}}.
\end{equation}

At state (C), where $V=V_{2}$, as light intensity increases so that phase transition to the isotropic state is induced, the order parameter reduces and capacitance decreases. Then the total energy function takes the form
\begin{equation}\label{NLC:eq:Wi}
\begin{split}
&\mathcal{W}_{i}(\lambda)
=\frac{\mu^{(1)}}{2}\left[\left(\frac{\mu^{(1)}/\mu^{(2)}+a^{1/3}/a_{0}^{1/3}}{\mu^{(1)}/\mu^{(2)}+a_{0}^{2/3}/a^{2/3}}\right)^{1/2}\lambda^{-1}+\left(\frac{\mu^{(1)}/\mu^{(2)}+a^{1/3}/a_{0}^{1/3}}{\mu^{(1)}/\mu^{(2)}+a_{0}^{2/3}/a^{2/3}}\right)^{-1/2}\lambda^{-1}+\lambda^{2}\right]\\
&+\frac{\mu^{(2)}}{2}\frac{a^{1/3}}{a_{0}^{1/3}}\left[\left(\frac{\mu^{(1)}/\mu^{(2)}+a^{1/3}/a_{0}^{1/3}}{\mu^{(1)}/\mu^{(2)}+a_{0}^{2/3}/a^{2/3}}\right)^{1/2}\lambda^{-1}a^{-1}a_{0}+\left(\frac{\mu^{(1)}/\mu^{(2)}+a^{1/3}/a_{0}^{1/3}}{\mu^{(1)}/\mu^{(2)}+a_{0}^{2/3}/a^{2/3}}\right)^{-1/2}\lambda^{-1}+\lambda^{2}\right]\\
&-\frac{\mu}{2}\varepsilon_{2}\varepsilon_{1}^{-1}V_{2}^2V_{1}^{-2}v\lambda^{-2}.
\end{split}
\end{equation}
Solving for $\lambda=\lambda_{i}$ and $v=v_{i}$ the  following system of equations,
\begin{equation}
\frac{\partial \mathcal{W}_{i}}{\partial\lambda}=0,\qquad \frac{\partial^{2} \mathcal{W}_{i}}{\partial\lambda^{2}}=0,
\end{equation}
then yields
\begin{equation}\label{NLC:eq:lambdai}
\lambda_{i}=\frac{1}{4^{1/3}}\frac{\left(\mu^{(1)}/\mu^{(2)}+a_{0}^{2/3}/a^{2/3}\right)^{1/6}}{\left(\mu^{(1)}/\mu^{(2)}+a^{1/3}/a_{0}^{1/3}\right)^{1/6}}
\end{equation}
and
\begin{equation}\label{NLC:eq:vi}
\begin{split}
v_{i}&=3\lambda_{i}^4\frac{\varepsilon_{1}}{\varepsilon_{2}}\frac{V_{1}^2}{V_{2}^2}\frac{\left(\mu^{(1)}/\mu^{(2)}+a^{1/3}/a_{0}^{1/3}\right)}{\mu^{(1)}/\mu^{(2)}+1}\\
&=\frac{3}{4^{4/3}}\frac{\varepsilon_{1}}{\varepsilon_{2}}\frac{V_{1}^2}{V_{2}^2}\frac{\left(\mu^{(1)}/\mu^{(2)}+a^{1/3}/a_{0}^{1/3}\right)^{1/3}\left(\mu^{(1)}/\mu^{(2)}+a_{0}^{2/3}/a^{2/3}\right)^{2/3}}{\mu^{(1)}/\mu^{(2)}+1}.
\end{split}
\end{equation}

From equations \eqref{NLC:eq:xi:CnCi} and \eqref{NLC:eq:C} we derive the capacitance ratio 
\begin{equation}\label{NLC:eq:xi:lambdani}
\xi=\frac{C_{n}}{C_{i}}=\frac{\varepsilon_{1}\lambda_{i}^2}{\varepsilon_{2}\lambda_{n}^2},
\end{equation}
and from equations \eqref{NLC:eq:Wm:xi}, \eqref{NLC:eq:Vm} and \eqref{NLC:eq:v} we obtain
\begin{equation}\label{NLC:eq:Wm}
W_{m}(\xi)=C_{n}V_{1}^2\frac{(\xi-1)^2}{4\xi}=\mu\Omega\frac{v(\xi-1)^2}{4\xi\lambda_{n}^2}
\end{equation}
as the optimal generated output per cycle.

%%%%%%%%%%%%%%%%%%%%%%%%%%%%%%%%%%%%
\subsection{The effect of pre-stretching perpendicular to the director}

Next, we consider the LCE dielectric to be pre-stretched in the second direction, i.e., perpendicular to the director, at state (A),  with a prescribed stretch ratio $\lambda_{0}>1$ \cite{Corbett:2009a:CW,Corbett:2009b:CW,Hiscock:2011:HWPM,Kofod:2008}. As pre-stretching increases the area of the dielectric and reduces the distance between plates, the amount of charge that can be taken from the battery increases. However, two types of instability may occur in this case, namely shear striping or wrinkling \cite{Corbett:2009a:CW,Corbett:2009b:CW}. The formation of shear stripes caused by director rotation in elongated nematic LCEs is well understood \cite{Finkelmann:1997:FKTW,Higaki:2013:HTU,Kundler:1995:KF,Kundler:1998:KF,Petelin:2009:PC,Petelin:2010:PC,Talroze:1999:etal,Zubarev:1999:etal} and has been modelled extensively \cite{Carlson:2002:CFS,Conti:2002a:CdSD,DeSimone:2000:dSD,DeSimone:2009:dST,Fried:2004:FS,Fried:2005:FS,Fried:2006:FS,Goriely:2021:GM,Kundler:1995:KF,Mihai:2022,Mihai:2020a:MG,Mihai:2021a:MG,Mihai:2023:MRGMG}. Wrinkling in compressed LCEs was examined theoretically in \cite{Goriely:2021:GM,Krieger:2019:KD,Soni:2016:SPP}. In pre-stretched LCEs, wrinkles can form if the voltage is too high, due to the so-called \textit{electrostrictive effect} observed when charging the electrodes. In this case, the applied Maxwell stress causes contraction in the field direction and elongation in the perpendicular directions. Here, we assume that the input voltage is below but close to the critical magnitude causing wrinkling, and that any reorientation of the nematic director that might occur is reverted  (see also Appendix~A). 

At the state (A), where $V=V_{1}$ and $a=a_{0}$, the energy function given by \eqref{NLC:eq:W:lambda} becomes
\begin{equation}
\mathcal{W}_{n}(\lambda,\lambda_{0})=\frac{\mu}{2}\left(\lambda^{-2}\lambda_{0}^{-2}+\lambda_{0}^{2}+\lambda^2\right)
-\frac{\mu}{2}v\lambda^{-2}.
\end{equation}
Solving for $\lambda=\lambda_{n}$ the equation
\begin{equation}
\frac{\partial \mathcal{W}_{n}}{\partial\lambda}=0
\end{equation}
yields
\begin{equation}\label{NLC:eq:lambdan:perp}
\lambda_{n}=\left(\lambda_{0}^{-2}-v\right)^{1/4}.
\end{equation}
In addition, solving for $v=v_{1}$ the following system of equations,
\begin{equation}
\frac{\partial \mathcal{W}_{n}}{\partial\lambda}=0,\qquad \frac{\partial\mathcal{W}_{n}}{\partial\lambda_{0}}=0,
\end{equation}
produces the wrinkling voltage
\begin{equation}\label{NLC:eq:v1:pre}
v_{1}=\lambda_{0}^{-2}-\lambda_{0}^{-8}.
\end{equation}
Note that wrinkling occurs when the stress in the second (pre-stretched) direction becomes zero, i.e., $P_{2}= \partial\mathcal{W}_{n}/\partial\lambda_{0}=0$.

At state (C), where $V=V_{2}$, the energy function is 
\begin{equation}
\begin{split}
\mathcal{W}_{i}(\lambda,\lambda_{0})
&=\frac{\mu^{(1)}}{2}\left(\lambda^{-2}\lambda_{0}^{-2}+\lambda_{0}^{2}+\lambda^2\right)
+\frac{\mu^{(2)}}{2}a^{1/3}a_{0}^{-1/3}\left(\lambda^{-2}\lambda_{0}^{-2}a^{-1}a_{0}+\lambda_{0}^{2}+\lambda^2\right)\\
&-\frac{\mu}{2}\varepsilon_{2}\varepsilon_{1}^{-1}V_{2}^2V_{1}^{-2}v\lambda^{-2}.
\end{split}
\end{equation}
Solving for $\lambda=\lambda_{i}$ the equation
\begin{equation}
\frac{\partial \mathcal{W}_{i}}{\partial\lambda}=0
\end{equation}
gives
\begin{equation}\label{NLC:eq:lambdai:perp}
\lambda_{i}=\left[\frac{\lambda_{0}^{-2}\left(\mu^{(1)}/\mu^{(2)}+a_{0}^{2/3}/a^{2/3}\right)-\left(\mu^{(1)}/\mu^{(2)}+1\right) \varepsilon_{2}\varepsilon_{1}^{-1}V_{2}^2V_{1}^{-2}v}{\mu^{(1)}/\mu^{(2)}+a^{1/3}/a_{0}^{1/3}}\right]^{1/4}.
\end{equation}
Then solving for $v=v_{2}$ the system of equations
\begin{equation}
\frac{\partial \mathcal{W}_{i}}{\partial\lambda}=0,\qquad \frac{\partial\mathcal{W}_{i}}{\partial\lambda_{0}}=0
\end{equation}
provides the wrinkling voltage
\begin{equation}\label{NLC:eq:v2:perp}
\begin{split}
v_{2}&=\frac{\varepsilon_{1}}{\varepsilon_{2}}\frac{V_{1}^2}{V_{2}^2}\left\{\frac{\mu^{(1)}/\mu^{(2)}}{\mu^{(1)}/\mu^{(2)}+1}\left[\lambda_{0}^{-2}-\lambda_{0}^{-8}\left(\frac{\mu^{(1)}/\mu^{(2)}+a_{0}^{2/3}/a^{2/3}}{\mu^{(1)}/\mu^{(2)}+a^{1/3}/a_{0}^{1/3}}\right)^2\right]\right.\\
&+\left.\frac{a^{1/3}/a_{0}^{1/3}}{\mu^{(1)}/\mu^{(2)}+1}\left[\lambda_{0}^{-2}a_{0}/a-\lambda_{0}^{-8}\left(\frac{\mu^{(1)}/\mu^{(2)}+a_{0}^{2/3}/a^{2/3}}{\mu^{(1)}/\mu^{(2)}+a^{1/3}/a_{0}^{1/3}}\right)^2\right]\right\}.
\end{split}
\end{equation}
In this case also, wrinkling is attained when the stress in the second direction is equal to zero, i.e., $P_{2}=\partial\mathcal{W}_{i}/\partial\lambda_{0}=0$.

From equations \eqref{NLC:eq:xi:CnCi}, \eqref{NLC:eq:xi:lambdani}, \eqref{NLC:eq:lambdan:perp}, and \eqref{NLC:eq:lambdai:perp}, we obtain the operating voltage
\begin{equation}\label{NLC:eq:v:perp}
v=\lambda_{0}^{-2}\frac{\xi^2\left(\mu^{(1)}/\mu^{(2)}+a^{1/3}/a_{0}^{1/3}\right)-\varepsilon_{1}^2\varepsilon_{2}^{-2}\left(\mu^{(1)}/\mu^{(2)}+a_{0}^{2/3}/a^{2/3}\right)}{\xi^2\left(\mu^{(1)}/\mu^{(2)}+a^{1/3}/a_{0}^{1/3}\right)-\varepsilon_{1}\varepsilon_{2}^{-1}\left(\mu^{(1)}/\mu^{(2)}+1\right)\left(\xi+1\right)^2/4},
\end{equation}
which is a nonlinear function of the pre-stretch ratio $\lambda_{0}$ and the capacitance ratio $\xi$. 

%%%%%%%%%%%%%%%%%%%%%%%%%%%%%%%%%%%%
\subsection{The effect of pre-stretching parallel to the director}

We also consider the case when the LCE is pre-stretched parallel to the nematic director at state (A), with a prescribed stretch ratio $\lambda_{0}>1$. In this case, at the state $(A)$, where $V=V_{1}$ and $a=a_{0}$, the energy function given by equation \eqref{NLC:eq:W:lambda} becomes
\begin{equation}
\mathcal{W}_{n}(\lambda,\lambda_{0})=\frac{\mu}{2}\left(\lambda_{0}^{2}+\lambda^{-2}\lambda_{0}^{-2}+\lambda^2\right)
-\frac{\mu}{2}v\lambda^{-2}.
\end{equation}
As before, solving for $\lambda=\lambda_{n}$ the equation
\begin{equation}
\frac{\partial \mathcal{W}_{n}}{\partial\lambda}=0
\end{equation}
gives
\begin{equation}\label{NLC:eq:lambdan:par}
\lambda_{n}=\left(\lambda_{0}^{-2}-v\right)^{1/4}.
\end{equation}
Then solving for $v=v_{1}$ the following system of equations,
\begin{equation}
\frac{\partial \mathcal{W}_{n}}{\partial\lambda}=0,\qquad \frac{\partial\mathcal{W}_{n}}{\partial\lambda_{0}}=0,
\end{equation}
yields the same wrinkling voltage as in equation \eqref{NLC:eq:v1:pre}.

At state (C),  where $V=V_{2}$, the energy function is equal to
\begin{equation}
\begin{split}
\mathcal{W}_{i}(\lambda,\lambda_{0})
&=\frac{\mu^{(1)}}{2}\left(\lambda_{0}^{2}+\lambda^{-2}\lambda_{0}^{-2}+\lambda^2\right)
+\frac{\mu^{(2)}}{2}a^{1/3}a_{0}^{-1/3}\left(\lambda_{0}^{2}a^{-1}a_{0}+\lambda^{-2}\lambda_{0}^{-2}+\lambda^2\right)\\
&-\frac{\mu}{2}\varepsilon_{2}\varepsilon_{1}^{-1}V_{2}^2V_{1}^{-2}v\lambda^{-2}.
\end{split}
\end{equation}
Then solving for $\lambda=\lambda_{i}$ the equation
\begin{equation}
\frac{\partial \mathcal{W}_{i}}{\partial\lambda}=0
\end{equation}
gives
\begin{equation}\label{NLC:eq:lambdai:par}
\lambda_{i}=\left[\frac{\lambda_{0}^{-2}\left(\mu^{(1)}/\mu^{(2)}+a^{1/3}/a_{0}^{1/3}\right)-\left(\mu^{(1)}/\mu^{(2)}+1\right) \varepsilon_{2}\varepsilon_{1}^{-1}V_{2}^2V_{1}^{-2}v}{\mu^{(1)}/\mu^{(2)}+a^{1/3}/a_{0}^{1/3}}\right]^{1/4}.
\end{equation}
Since the LCE tends to contract in the pre-stretched direction while expanding in the thickness direction, there is no wrinkling.

By equations \eqref{NLC:eq:xi:CnCi}, \eqref{NLC:eq:xi:lambdani}, \eqref{NLC:eq:lambdan:par}, and \eqref{NLC:eq:lambdai:par}, we obtain
\begin{equation}\label{NLC:eq:v:par}
v=\lambda_{0}^{-2}\frac{\xi^2\left(\mu^{(1)}/\mu^{(2)}+a^{1/3}/a_{0}^{1/3}\right)-\varepsilon_{1}^2\varepsilon_{2}^{-2}\left(\mu^{(1)}/\mu^{(2)}+a^{1/3}/a_{0}^{1/3}\right)}{\xi^2\left(\mu^{(1)}/\mu^{(2)}+a^{1/3}/a_{0}^{1/3}\right)-\varepsilon_{1}\varepsilon_{2}^{-1}\left(\mu^{(1)}/\mu^{(2)}+1\right)\left(\xi+1\right)^2/4},
\end{equation}
as a nonlinear function of $\lambda_{0}$ and $\xi$. 

%%%%%%%%%%%%%%%%%%%%%%%%%%%%%%%%%%%%%%%%%%%%%%%%%%%%%%%%%%%%
%%%%%%%%%%%%%%%%%%%%  NEW SECTION   %%%%%%%%%%%%%%%%%%%%%%%%
%%%%%%%%%%%%%%%%%%%%%%%%%%%%%%%%%%%%%%%%%%%%%%%%%%%%%%%%%%%%
\section{Numerical results}\label{NLC:sec:results:cp}

\begin{figure}[htbp] 
	\begin{center}
		(a)\includegraphics[width=0.45\textwidth]{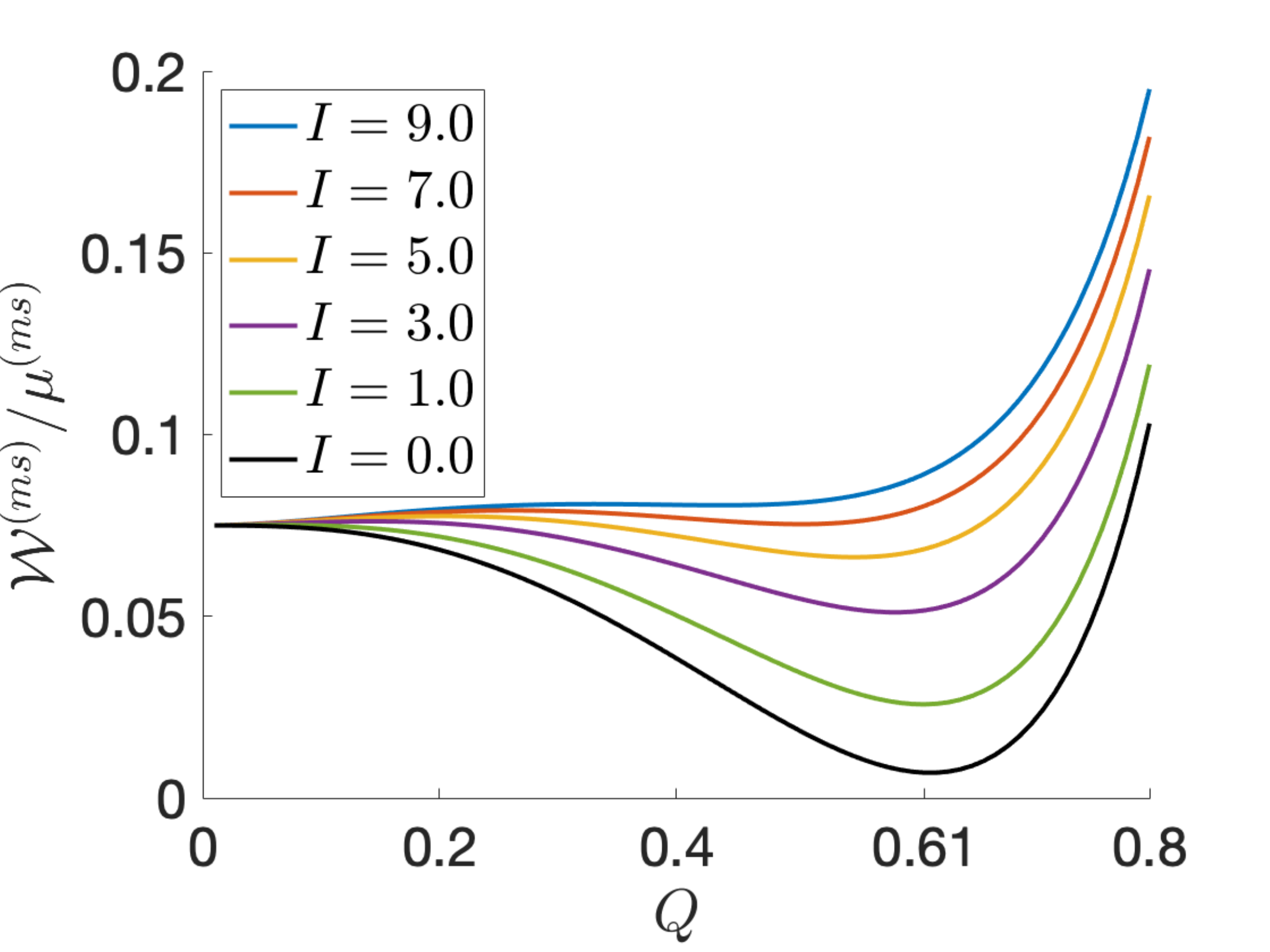}
		(b)\includegraphics[width=0.45\textwidth]{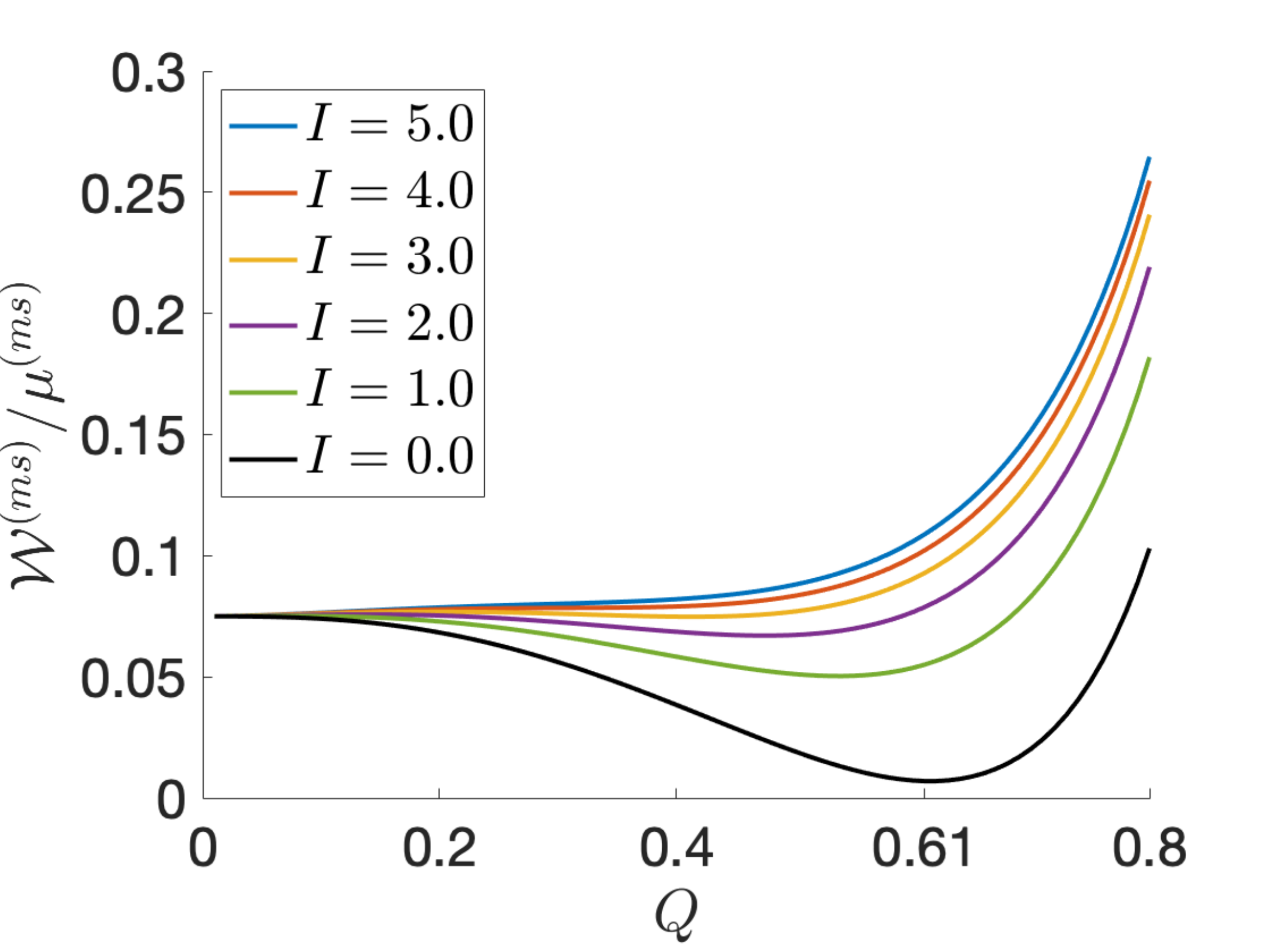}
		\caption{The modified Maier-Saupe mean field energy given by equation~\eqref{NLC:eq:Wms:cp}  as a function of the order parameter $Q$ when there is no light ($I=0$) or the light is: (a) polarised or (b) unpolarised, with varying intensity $I>0$ and $\alpha=\pi/2$, $f=1/6$, $\tilde{J}=5$. As $I$ increases, there is a transition from the nematic phase to the isotropic phase. This transition occurs at higher values of $I$ when the light is polarised than when it is unpolarised. When $I=0$, the minimum energy is attained for $Q_{0}=0.61$.}\label{NLC:fig:Q0:light}
	\end{center}
\end{figure}

In this section, we present a set of numerical results to illustrate the performance of the theoretical model for the LCE-based charge pump. Following \cite{Hiscock:2011:HWPM}, we choose the shear modulus $\mu=10^{6}$ Pa and initial thickness $d_{0}=50\cdot 10^{-6}$ m for the LCE, and the dielectric constants $\varepsilon_{0\parallel}=10$ and $\varepsilon_{0\perp}=20$.  However, here, $\mu=\mu^{(1)}+\mu^{(2)}$ and the ratio $\mu^{(1)}/\mu^{(2)}$ can vary.
\begin{itemize}
	\item When heat is absorbed, we take the scalar uniaxial order parameters $Q_{0}=0.5$ and $Q=0.02$, corresponding to the nematic and isotropic states, respectively \cite{Hiscock:2011:HWPM};
	\item When light is absorbed, we set $\alpha=\pi/2$, $f=1/6$, $\tilde{J}=5$ and  $\mu/\mu^{(ms)}=0.05$ \cite{Bai:2020:BB,Corbett:2006:CW,Corbett:2008:CW}. At the initial state where there is no light, $I=0$ and the optimal order parameter is $Q_{0}=0.61$, while at the isotropic state, $Q=0$  (see Figure~\ref{NLC:fig:Q0:light}). 
\end{itemize}
The following parameters can then calculated directly: $\varepsilon_{1}=\left[\varepsilon_{0\parallel}\left(1-Q_{0}\right)+\varepsilon_{0\perp}\left(2+Q_{0}\right)\right]/3=\bar{\varepsilon}+Q_{0}\left(\varepsilon_{0\perp}-\bar{\varepsilon}\right)$ and $\varepsilon_{2}=\bar{\varepsilon}+Q\left(\varepsilon_{0\perp}-\bar{\varepsilon}\right)\approx\bar{\varepsilon}$, where $\bar{\varepsilon}=\left(2\varepsilon_{0\perp}+\varepsilon_{0\parallel}\right)/3$.

When light is absorbed, the required (solar) input so that the LCE transitions from the nematic to the isotropic state is assumed $H_{light}\approx 10^{7}$ J/m$^{3}$ \cite{Hiscock:2011:HWPM}, while when the LCE absorbs heat, the energy needed is considered $H_{heat}\approx3\cdot 10^{6}$ J/m$^{3}$ \cite[Section~2.3]{Warner:2007:WT}. Then the efficiency of the system absorbing is given by the ratio between the generated output per cycle and the required input. For the three cases where the elastomer is not pre-stretched and when it is pre-stretched either perpendicular or parallel to the nematic director, this is summarised in Figure~\ref{NLC:fig:eff}.

%%%%%%%%%%%%%%%%%%%%%%%%%%%%%%%%%%%%
\subsection{Energy efficiency under natural deformation}

\begin{figure}[htbp] 
	\begin{center}
		(a) \includegraphics[width=0.45\textwidth]{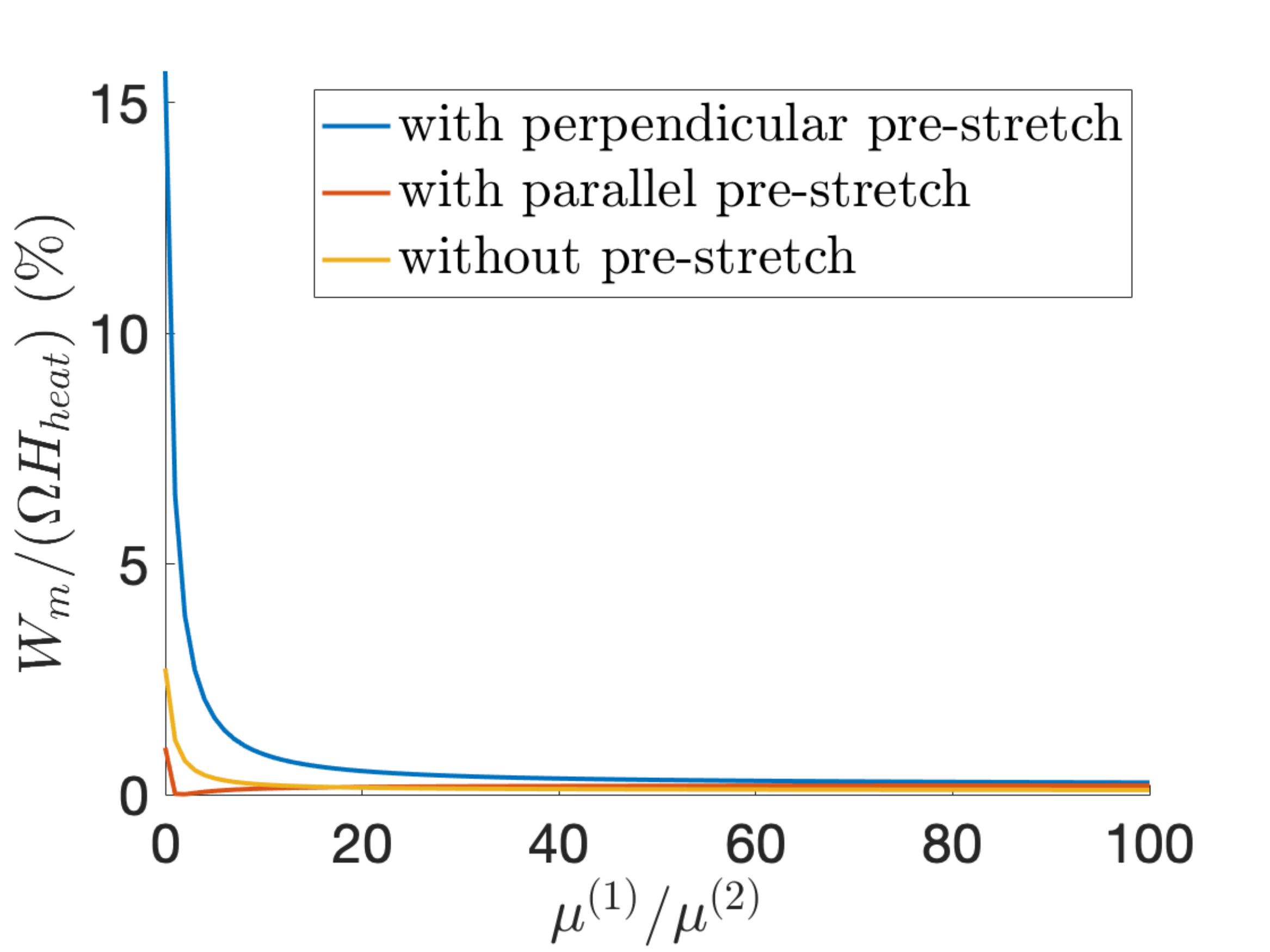}
		(b) \includegraphics[width=0.45\textwidth]{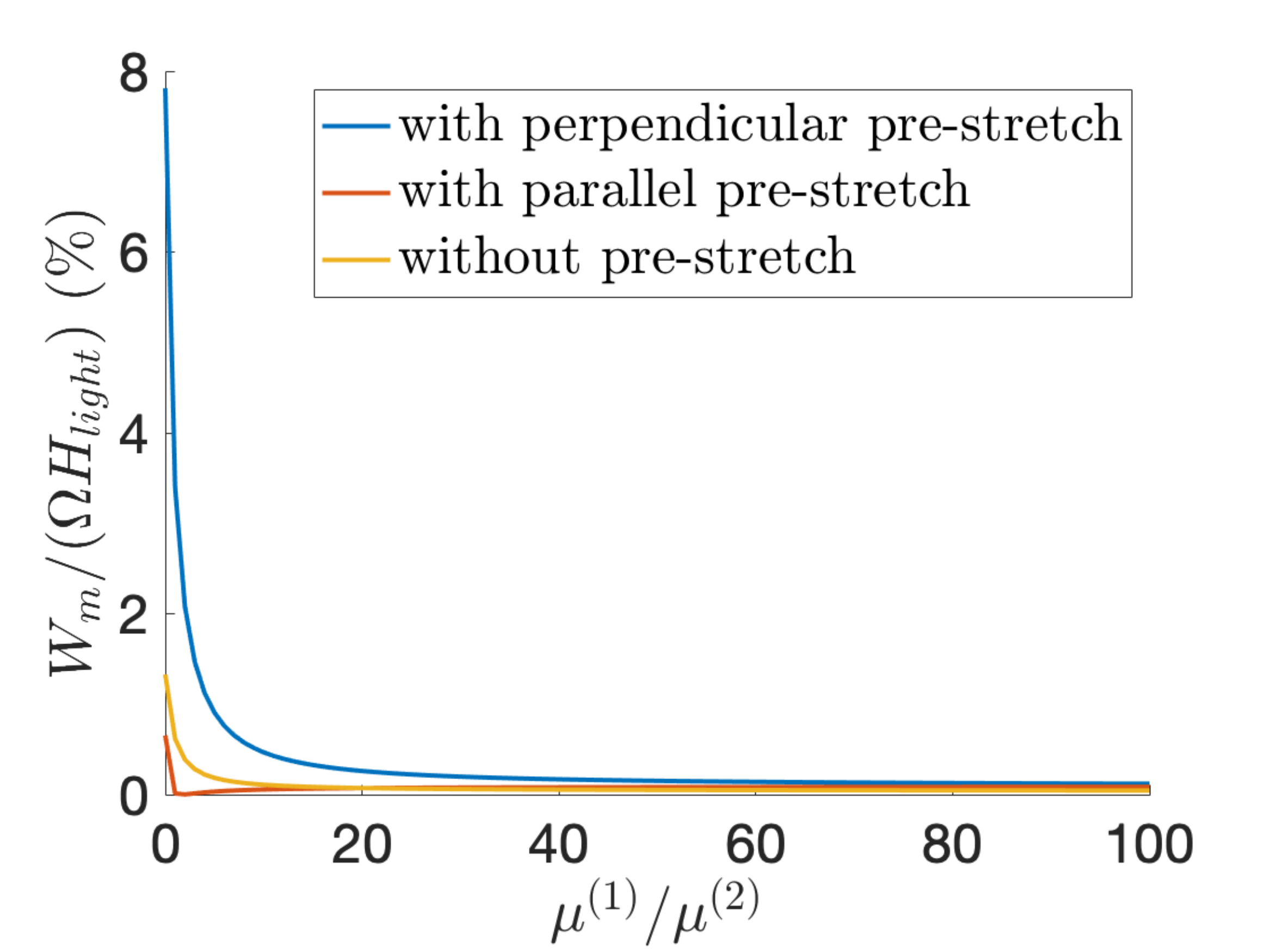}
		\caption{The efficiency bound as a function of the parameter ratio $\mu^{(1)}/\mu^{(2)}>0$ for the LCE dielectric absorbing: (a) heat or (b) light when there is no initial pre-stretch or when there is a pre-stretch either perpendicular or parallel to the nematic director, with ratio $\lambda_{0}=1.25$.}\label{NLC:fig:eff}
	\end{center}
\end{figure}

\begin{figure}[htbp] 
	\begin{center}
		(a) \includegraphics[width=0.45\textwidth]{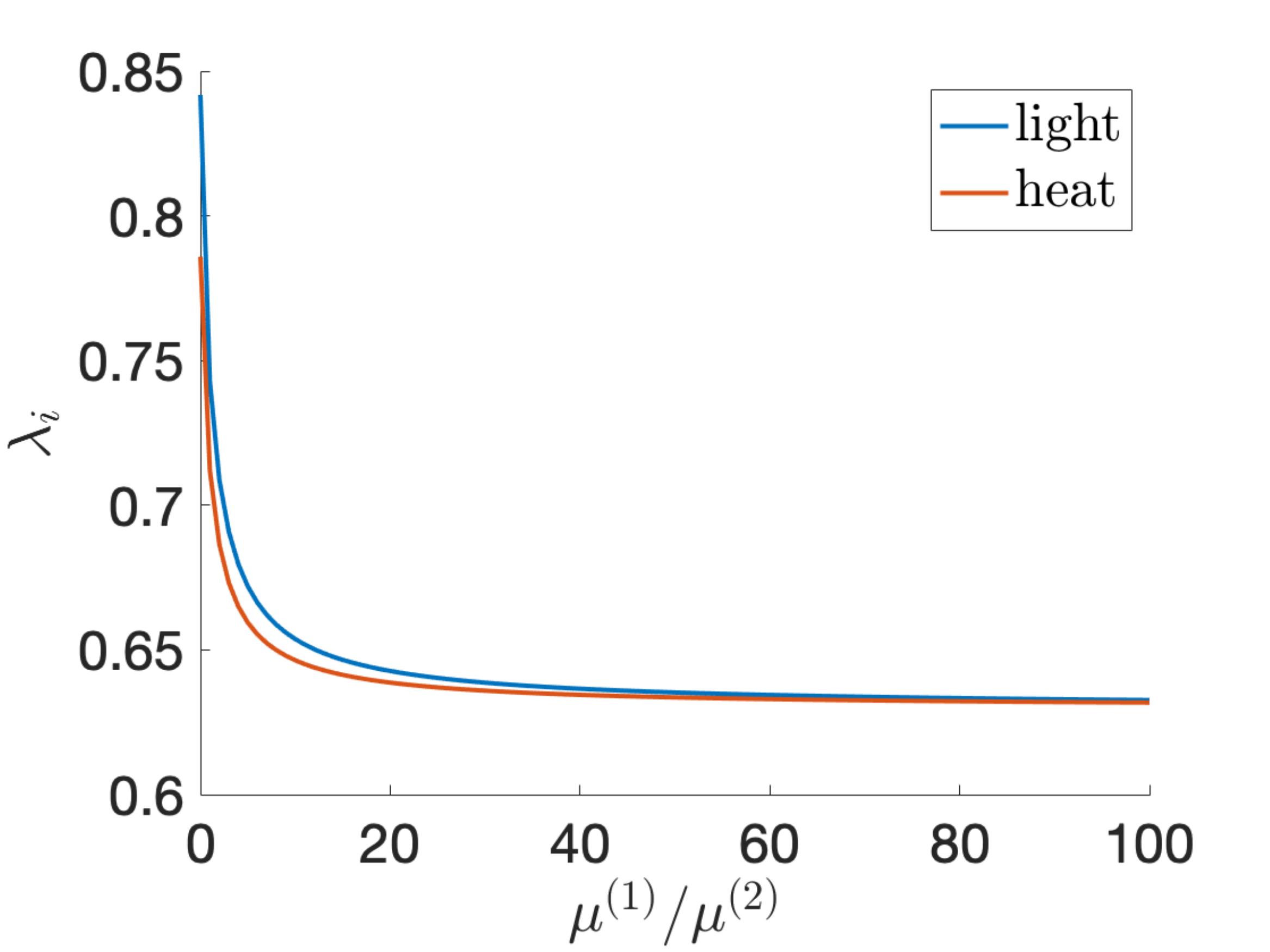}
		(b) \includegraphics[width=0.45\textwidth]{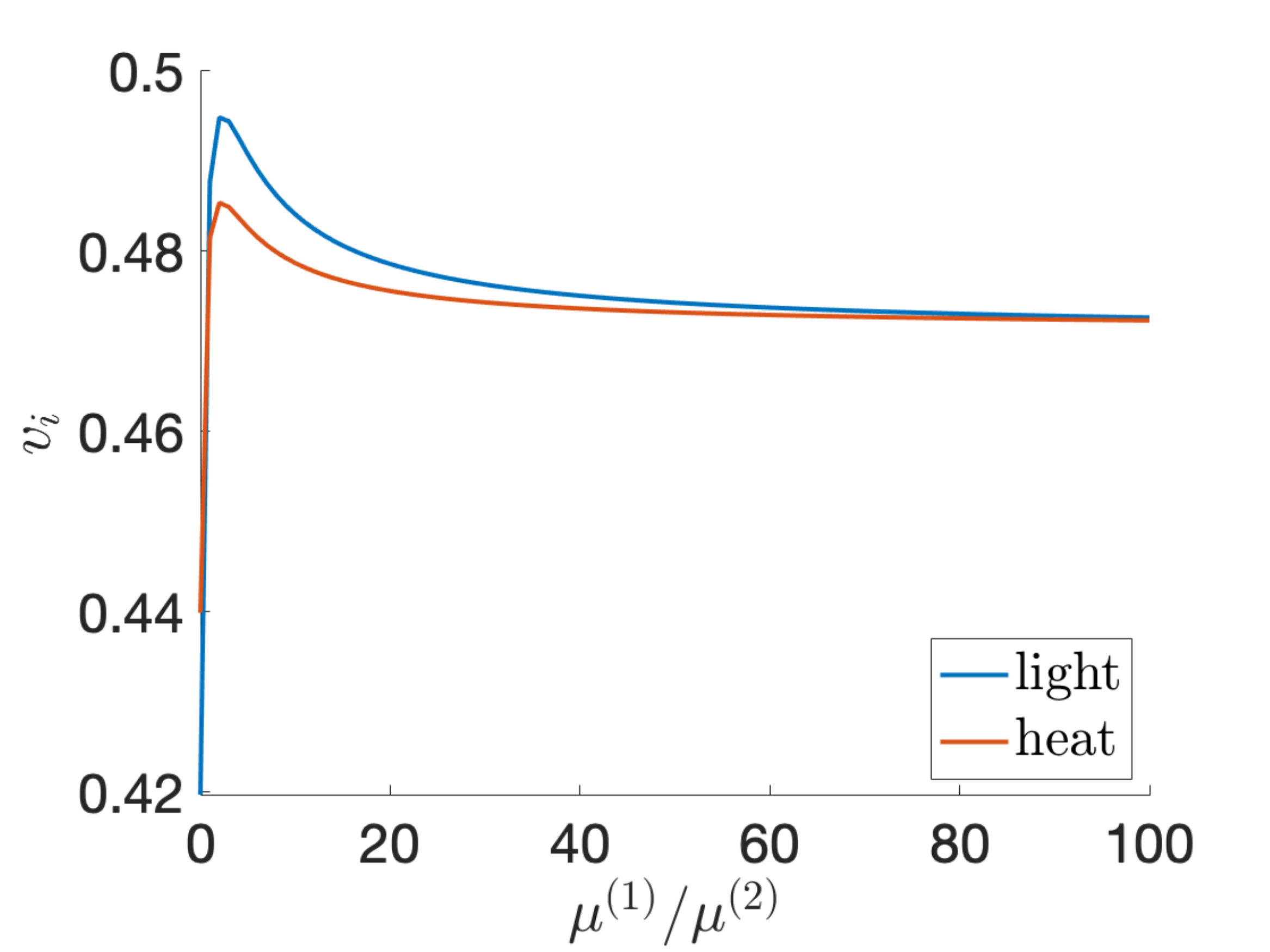}
		\caption{(a) The stretch ratio $\lambda_{i}$ given by equation \eqref{NLC:eq:lambdai} and (b)  the voltage $v_{i}$ satisfying equation \eqref{NLC:eq:vi} as functions of the parameter ratio $\mu^{(1)}/\mu^{(2)}$ for the LCE dielectric absorbing heat or light.}\label{NLC:fig:lambdaivi}
	\end{center}
\end{figure}

\begin{figure}[htbp] 
	\begin{center}
		(a) \includegraphics[width=0.45\textwidth]{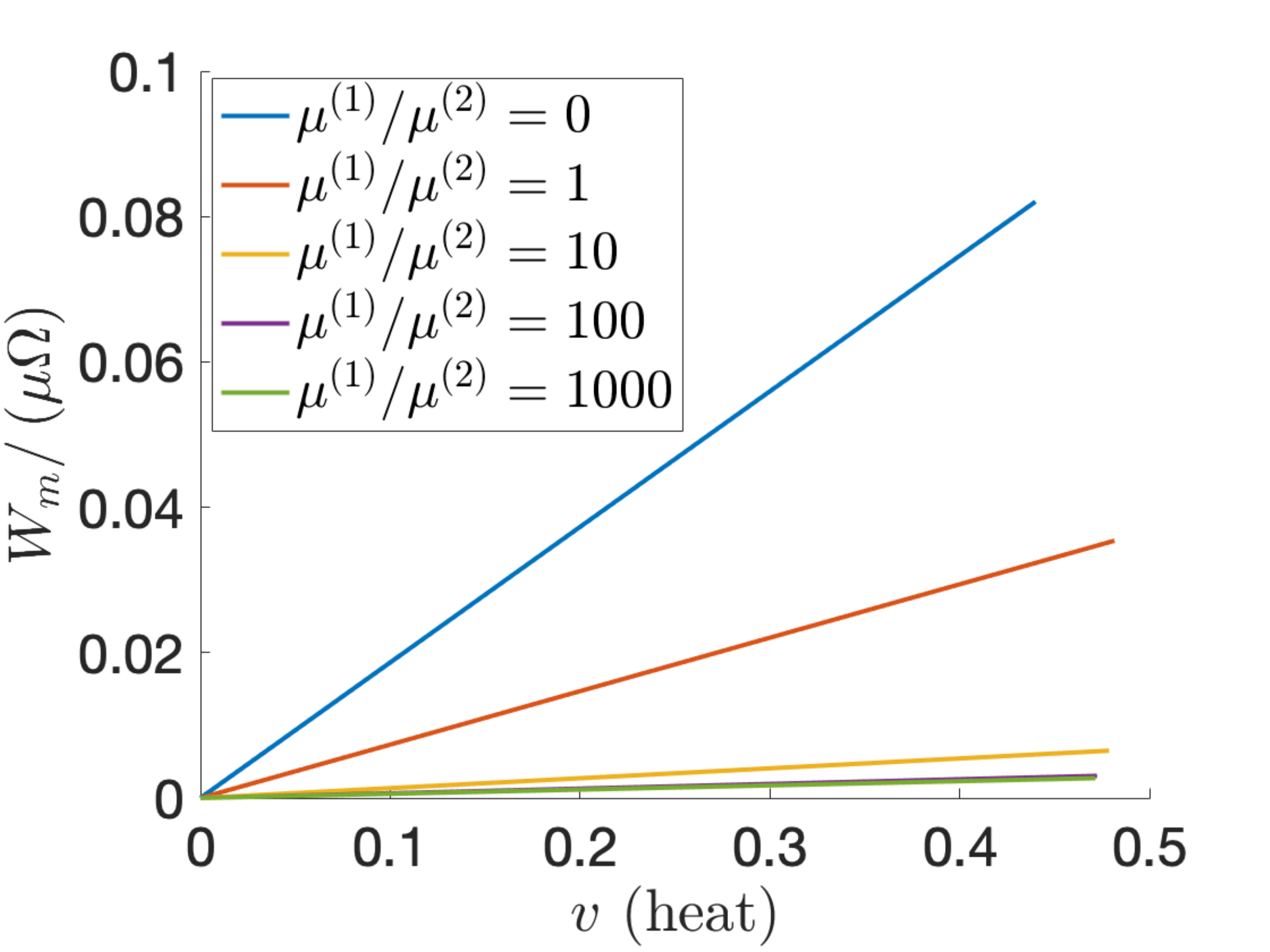}
		(b)  \includegraphics[width=0.45\textwidth]{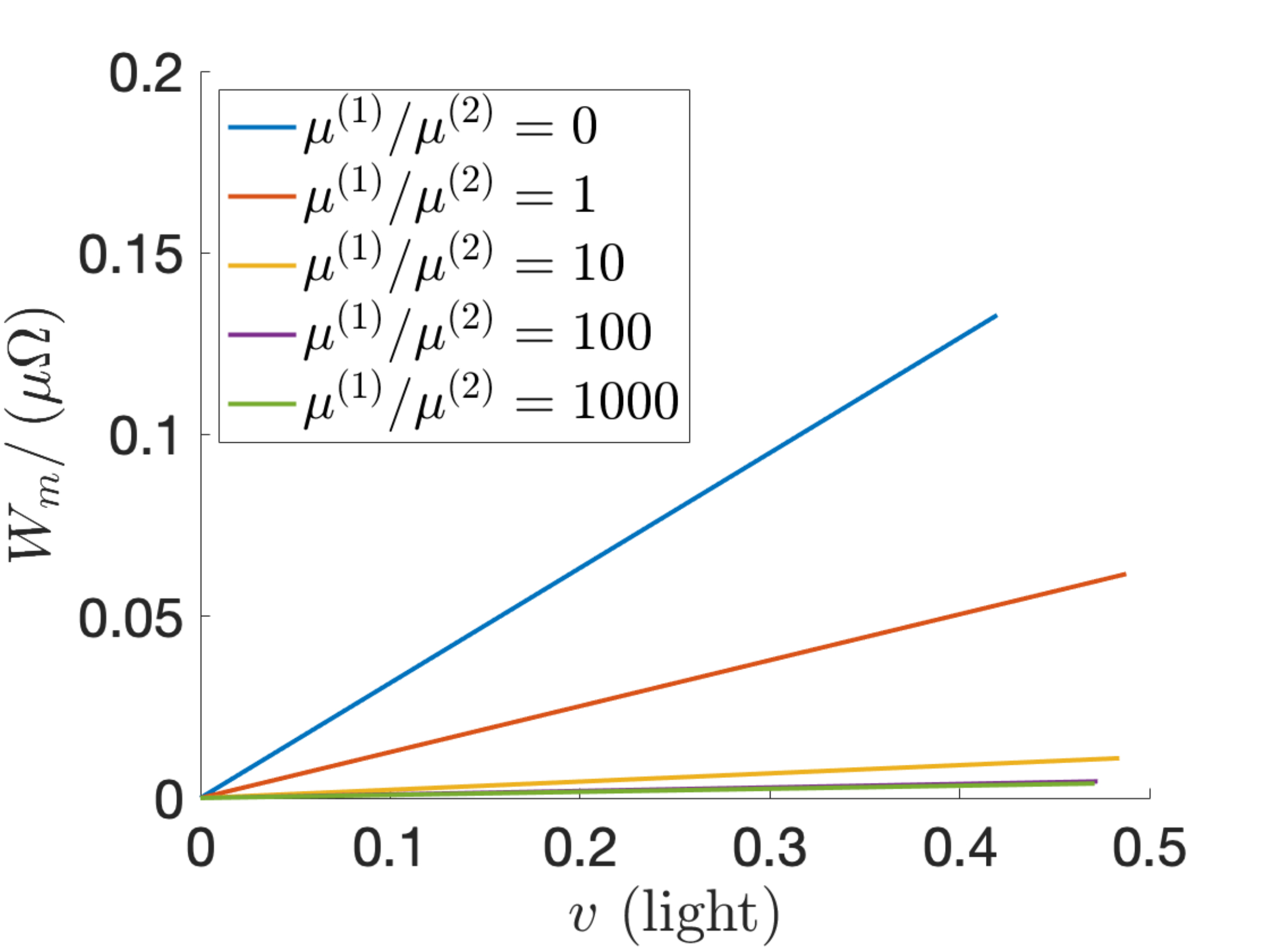}
		\caption{The scaled optimal generated output $W_{m}/(\mu\Omega)$ for $\mu^{(1)}/\mu^{(2)}\in\{0,1,10,100,1000\}$ when the LCE dielectric absorbs: (a) heat or (b) light.}\label{NLC:fig:Wm}
	\end{center}
\end{figure}

If the LCE is in its natural configuration at initial state (A), then $C=C_{n}$, $\varepsilon_{\perp}=\varepsilon_{1}$, and $\lambda=\lambda_{n}$, while at state (C), $C=C_{i}$, $\varepsilon_{\perp}=\varepsilon_{2}$, and $\lambda=\lambda_{i}$. The stretch ratio $\lambda_{i}$ and the corresponding operating voltage $v_{i}$, defined by equations \eqref{NLC:eq:lambdai} and \eqref{NLC:eq:vi}, respectively, are plotted as functions of the parameter ratio $\mu^{(1)}/\mu^{(2)}$ in Figure~\ref{NLC:fig:lambdaivi}. By varying this ratio, the maximum optimal output per unit volume is shown in Figure~\ref{NLC:fig:Wm}. For example, if $\mu^{(1)}/\mu^{(2)}=1$, then:
\begin{itemize}
	\item When the LCE absorbs heat, the maximum optimal output is equal to $W_{m}/\Omega\approx 3.5\cdot 10^4$ J/m$^{3}$ per cycle. The efficiency is $W_{m}/\left(H_{heat}\Omega\right)\approx0.035/3\approx 1.2\%$. The operating voltages are $V_{1}=\left[\mu v_{i}d_{0}^2/\left(\varepsilon_{1}\varepsilon_{0\perp}\right)\right]^{1/2}\approx 1.8$ kV and $V_{2}=V_{1}\left(\xi+1\right)/2\approx 2.2$ kV  (see also Appendix~B).
	\item When the LCE absorbs light, the maximum optimal output is $W_{m}/\Omega\approx 0.6\cdot 10^5$ J/m$^{3}$ per cycle and the efficiency is $W_{m}/\left(H_{light}\Omega\right)\approx 0.6\%$. The operating voltages are $V_{1}\approx 1.8$ kV and $V_{2}\approx 2.3$ kV.
\end{itemize}
For these two cases, Figure~\ref{NLC:fig:eff} shows that efficiency decreases as $\mu^{(1)}/\mu^{(2)}$ increases. Since $\mu^{(1)}/\mu^{(2)}=0$ corresponds to the neoclasical model for ideal LCEs, while $\mu^{(1)}/\mu^{(2)}\to\infty$ corresponds to the neo-Hookean model for rubber, this figure suggests that LCEs are more efficient than rubber in generating electricity.

%%%%%%%%%%%%%%%
%%%%%%%%%%%%%%%

%%%%%%%%%%%%%%%
%%%%%%%%%%%%%%%

%%%%%%%%%%%%%%%
%%%%%%%%%%%%%%%
%%%%%%%%%%%%%%%
%%%%%%%%%%%%%%%

%%%%%%%%%%%%%%%
\begin{figure}[htbp] 
	\begin{center}
		(a)\includegraphics[width=0.45\textwidth]{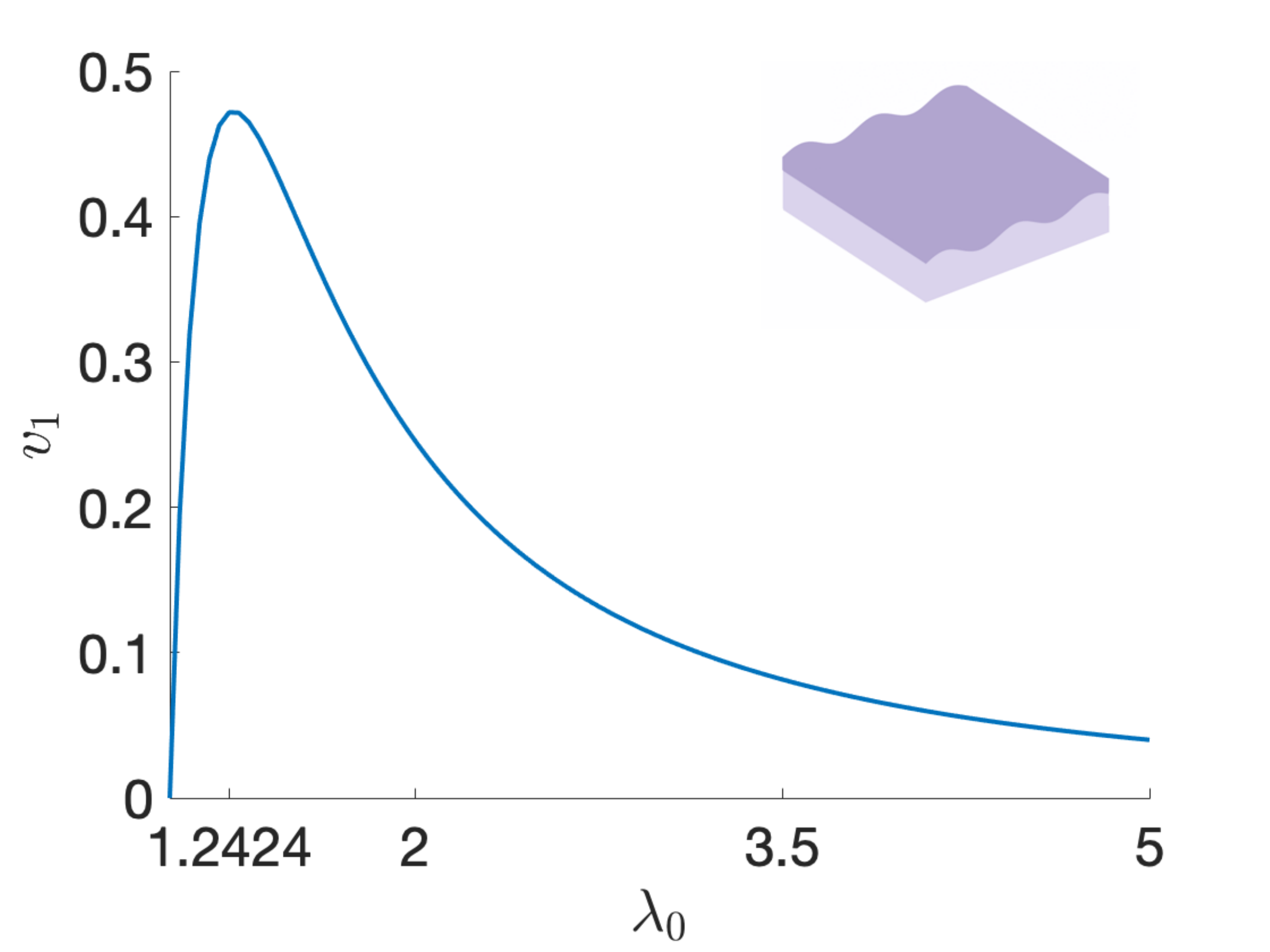}
		(b)\includegraphics[width=0.45\textwidth]{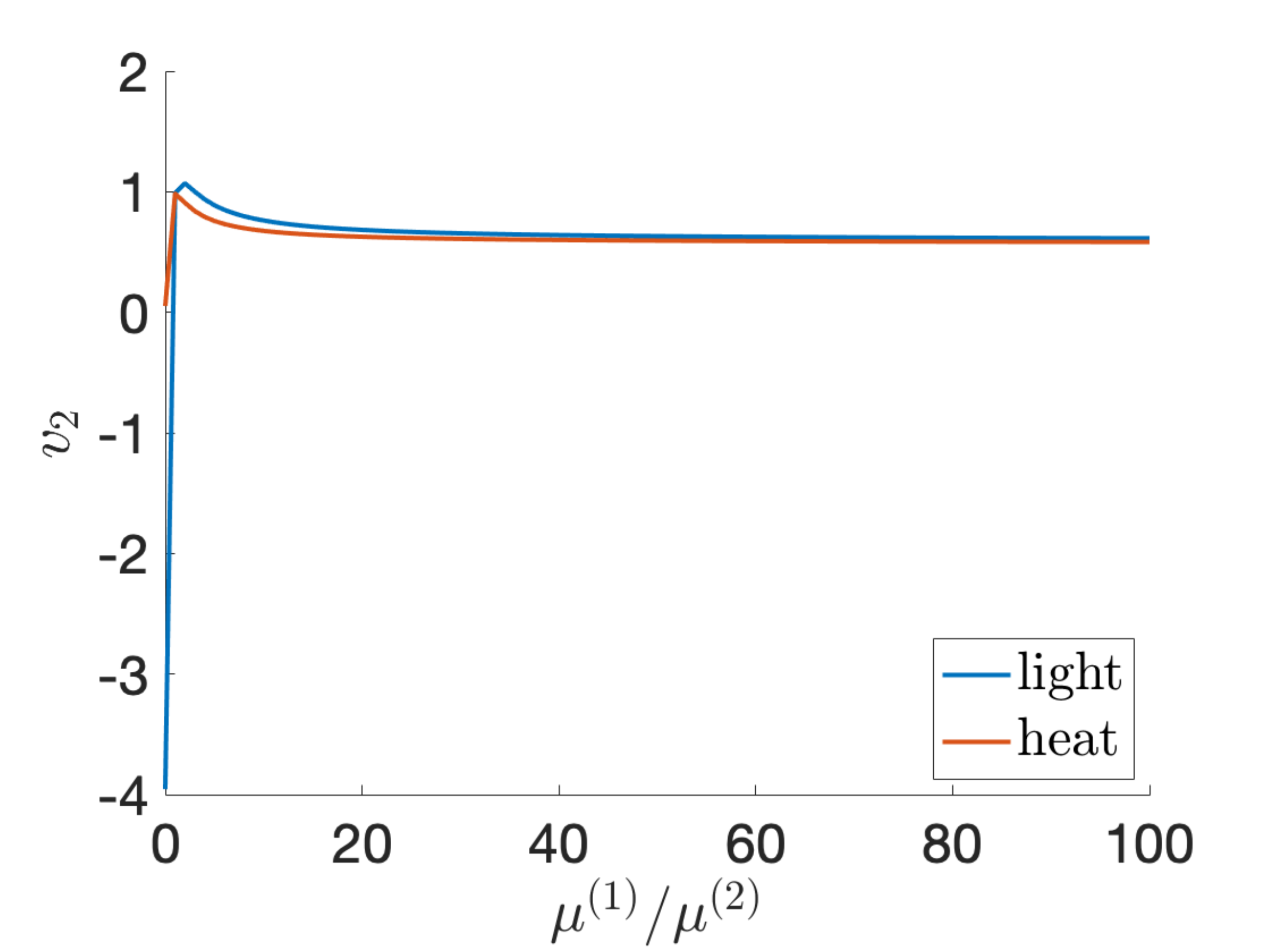}
		\caption{Wrinkling (a) input voltage $v_{1}$ given by equation \eqref{NLC:eq:v1:pre} as a function of the pre-stretch ratio $\lambda_{0}$ and  (b) output voltage $v_{2}$ given by equation \eqref{NLC:eq:v2:perp} as a function of  the parameter ratio $\mu^{(1)}/\mu^{(2)}$ when the LCE dielectric is pre-stretched perpendicular to the director with ratio $\lambda_{0}=1.25$ and absorbs heat or light. The maximum input wrinkling voltage is attained for $\lambda_{0}=1.2424$.}\label{NLC:fig:v1v2:pre}
	\end{center}
\end{figure}
%%%%%%%%%%%%%%%
%%%%%%%%%%%%%%%
\begin{figure}[htbp] 
	\begin{center}
		(a)\includegraphics[width=0.45\textwidth]{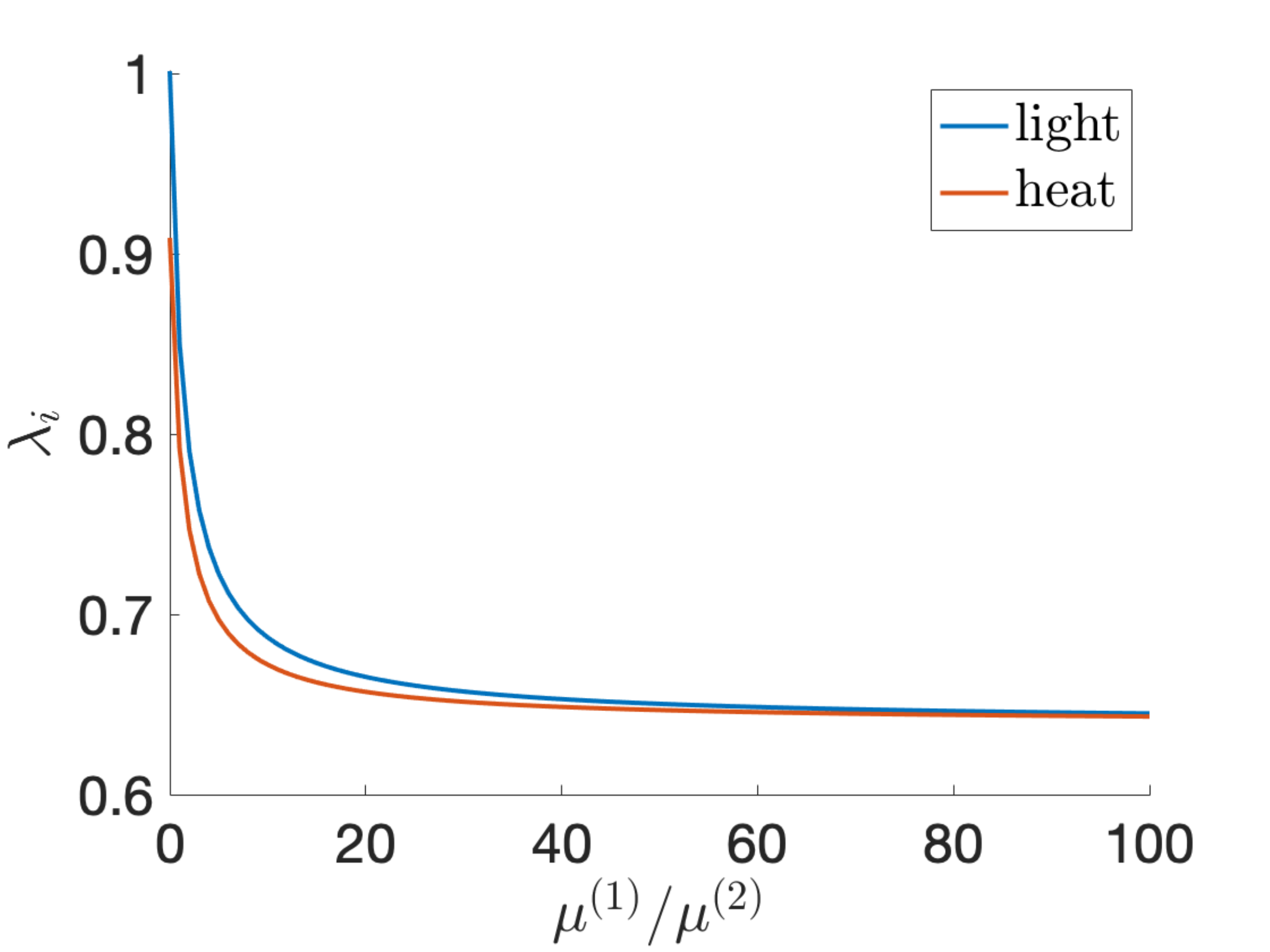}
		(b)\includegraphics[width=0.45\textwidth]{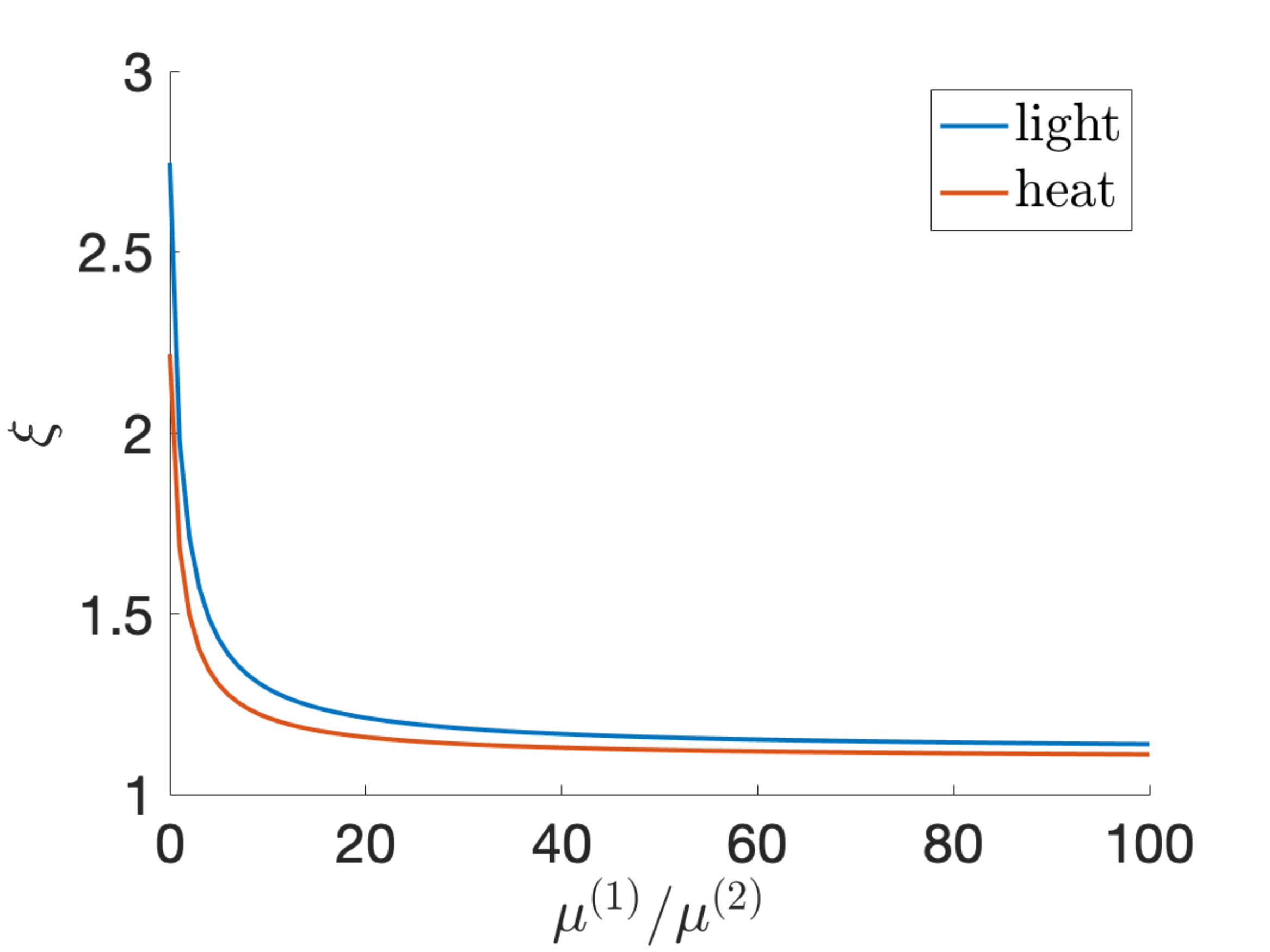}
		\caption{(a) The stretch ratio $\lambda_{i}$ given by equation \eqref{NLC:eq:lambdai:perp} and (b) the capacitance ratio $\xi$ satisfying equation \eqref{NLC:eq:v:perp} as functions of the parameter ratio $\mu^{(1)}/\mu^{(2)}$ when the LCE dielectric is pre-stretched perpendicular to the director with ratio $\lambda_{0}=1.25$ and absorbs heat or light.}\label{NLC:fig:lambdaixi:perp}
	\end{center}
\end{figure}
%%%%%%%%%%%%%%%

%%%%%%%%%%%%%%%
\begin{figure}[htbp] 
	\begin{center}
		(a)\includegraphics[width=0.45\textwidth]{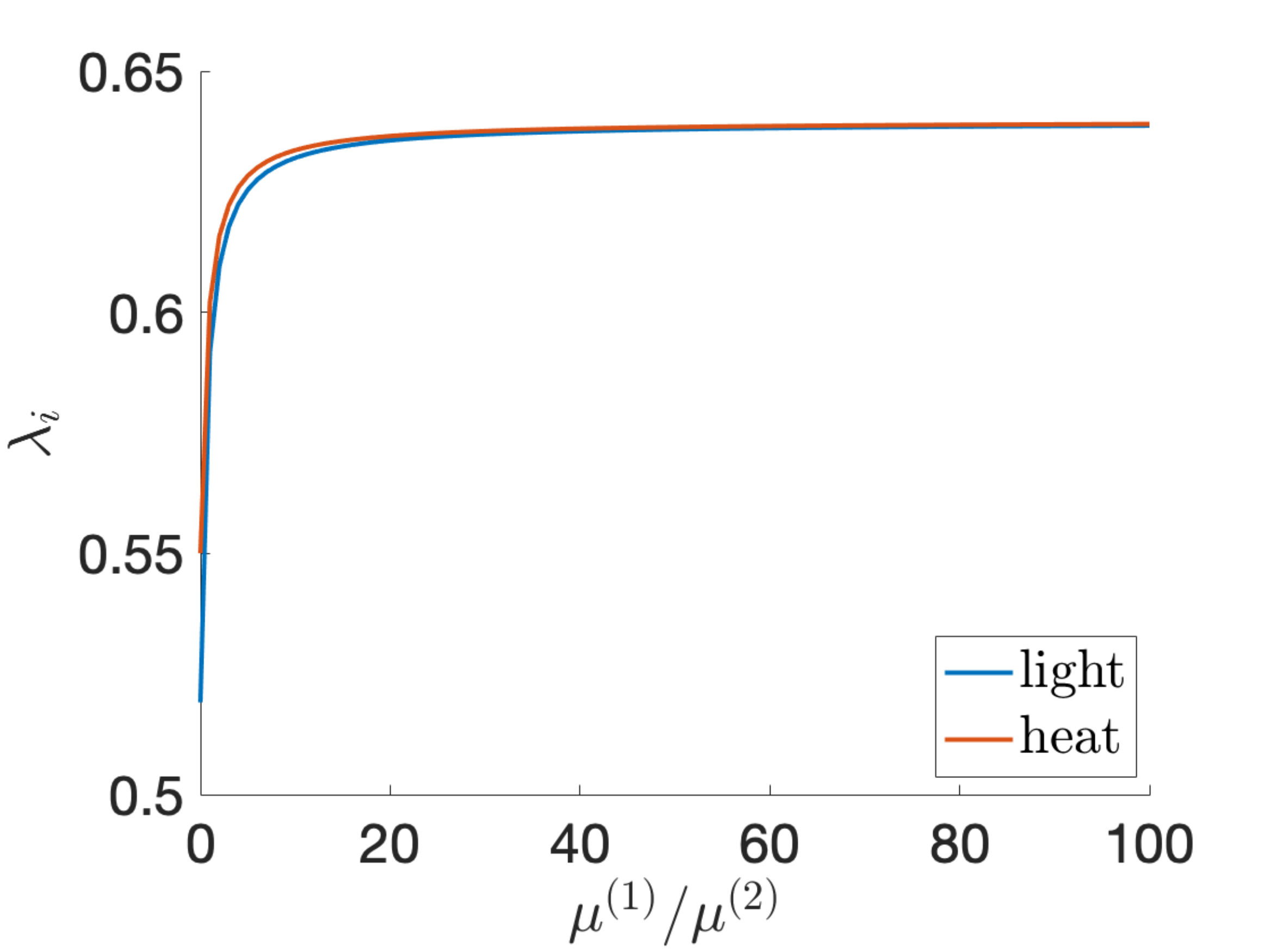}
		(b)\includegraphics[width=0.45\textwidth]{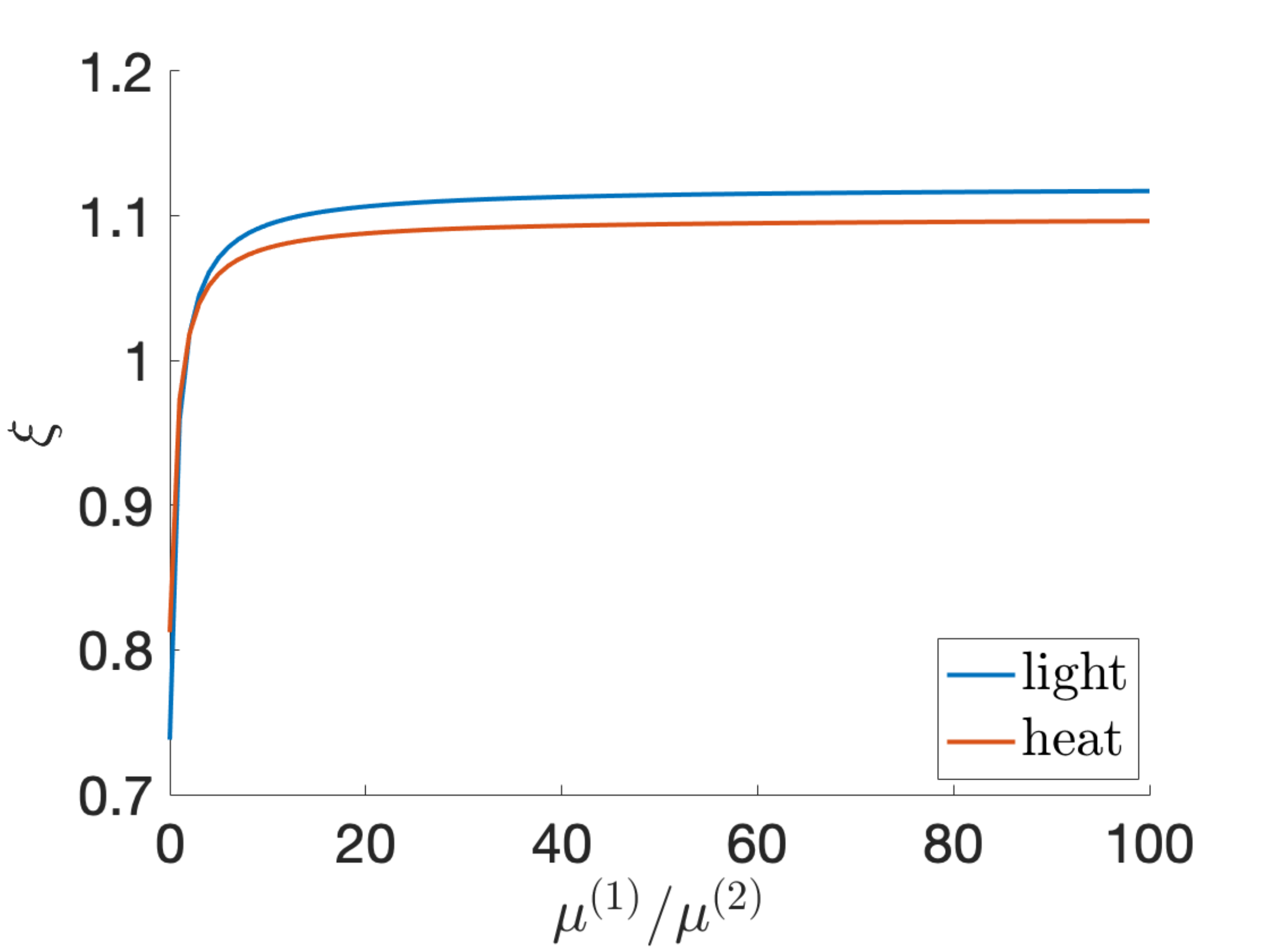}
		\caption{(a) The stretch ratio $\lambda_{i}$ given by equation \eqref{NLC:eq:lambdai:par} and (b) the capacitance ratio $\xi$ satisfying equation \eqref{NLC:eq:v:par} as functions of the parameter ratio $\mu^{(1)}/\mu^{(2)}$ when the LCE dielectric is pre-stretched parallel to the director with ratio $\lambda_{0}=1.25$ and absorbs heat or light.}\label{NLC:fig:lambdaixi:par}
	\end{center}
\end{figure}
%%%%%%%%%%%%%%%

%%%%%%%%%%%%%%%%%%%%%%%%%%%%%%%%%%%%
\subsection{Energy efficiency when pre-stretching perpendicular to the director}

When, at state (A), the LCE dielectric is pre-stretched perpendicular to the director, with initial stretch ratio $\lambda_{0}>1$, the input wrinkling voltage, given by equation \eqref{NLC:eq:v1:pre} is independent of $\mu^{(1)}/\mu^{(2)}$ and attains its maximum for $\lambda_{0}=1.2424$, while the output wrinkling voltage, given by equation \eqref{NLC:eq:v2:perp}, decreases as the parameter ratio $\mu^{(1)}/\mu^{(2)}$ increases. The input wrinkling voltage is plotted in Figure~\ref{NLC:fig:v1v2:pre}(a). To maximise the operating voltage, we choose an input close to the corresponding wrinkling voltage. 
For example, when $\lambda_{0}=1.25$ and $v=0.4$, if the parameter ratio $\mu^{(1)}/\mu^{(2)}$ varies, then the output wrinkling voltage $v_{2}$ given by equation \eqref{NLC:eq:v2:perp} is displayed in Figure~\ref{NLC:fig:v1v2:pre}(b), and the stretch ratio $\lambda_{i}$ given by equation \eqref{NLC:eq:lambdai:perp} and capacitance ratio $\xi$ satisfying equation \eqref{NLC:eq:v:perp} are plotted in Figure~\ref{NLC:fig:lambdaixi:perp}. In this case, Figure~\ref{NLC:fig:eff} suggests that efficiency decreases as $\mu^{(1)}/\mu^{(2)}$ increases. In particular, if $\mu^{(1)}/\mu^{(2)}=1$, then:
\begin{itemize}
	\item When the LCE absorbs heat, the maximum optimal output is equal to $W_{m}/\Omega 2\cdot 10^5$ J/m$^3$ per cycle. The efficiency is $W_{m}/\left(H_{heat}\Omega\right)=0.2/3\approx 6\%$. The operating voltages are $V_{1}\approx 1.8$ kV and $V_{2}\approx 2.4$ kV. 
	\item When the LCE absorbs light, the maximum optimal output is $W_{m}/\Omega=3.4\cdot 10^5$ J/m$^3$ per cycle. The efficiency is $W_{m}/\left(H_{light}\Omega\right)\approx 3.4\%$. The operating voltages are $V_{1}\approx 1.8$ kV and $V_{2}\approx 2.6$ kV. 
\end{itemize}
For the numerical values of the given parameters, the auxiliary results presented in Appendix A imply that director rotation can be ignored when the LCE dielectric is pre-stretched perpendicular to the director.

%%%%%%%%%%%%%%%%%%%%%%%%%%%%%%%%%%%%
\subsection{Energy efficiency when pre-stretching parallel to the director}

When the LCE dielectric is pre-stretched parallel to the director, the input wrinkling voltage is the same as that shown in Figure~\ref{NLC:fig:v1v2:pre}(a). Choosing again $\lambda_{0}=1.25$ and $v=0.4$, when the parameter ratio $\mu^{(1)}/\mu^{(2)}$ varies, the stretch ratio $\lambda_{i}$ given by equation \eqref{NLC:eq:lambdai:par} and capacitance ratio $\xi$ satisfying equation \eqref{NLC:eq:v:par} are represented in Figure~\ref{NLC:fig:lambdaixi:par}. We note that $\xi=C_{n}/C_{i}>1$ for $\mu^{(1)}/\mu^{(2)}>1.6$ when the LCE is heated and for $\mu^{(1)}/\mu^{(2)}>1.5$ when the LCE is illuminated. In this case, the efficiency shown in Figure~\ref{NLC:fig:eff} increases with $\mu^{(1)}/\mu^{(2)}$.

%%%%%%%%%%%%%%%%%%%%%%%%%%%%%%%%%%%%%%%%%%%%%%%%%%%%%%%%%%%%
%%%%%%%%%%%%%%%%%%%%  NEW SECTION   %%%%%%%%%%%%%%%%%%%%%%%%
%%%%%%%%%%%%%%%%%%%%%%%%%%%%%%%%%%%%%%%%%%%%%%%%%%%%%%%%%%%%
\section{Conclusion}\label{NLC:sec:conclusion:cp}

In this study, a theoretical model is developed for a charge pump with a parallel plate capacitor where the dielectric is made of LCE material that naturally responds to environmental changes such as heat or light. Specifically, heating or illuminating the LCE induces a transition from a nematic to an isotropic state. In addition, the geometry of the dielectric changes, and the contact area with the conducting plates and the distance between them are altered. In the charge pump electrical circuit, at the beginning of a reversible cycle of heating and cooling or illumination and absence of light, first the dielectric is assumed to be in a relaxed natural state, then it is pre-stretched so that the capacitance increases by increasing the contact area with the plates and decreasing the distance between them. The LCE is described by a composite strain-energy function which, when taking its constitutive parameters to their limiting values, can be reduced to either the purely elastic neo-Hookean model or the neoclassical model for ideal nematic elastomers. From the above analysis, we infer that: (i) LCE is more efficient than rubber when used as dielectric in a parallel plate capacitor; (ii) When the dielectric is pre-stretched perpendicular to the director at the initial state of the proposed cycle, the capacitor becomes more effective in raising the voltage supplied by the source battery.

To make these results analytically tractable, in the proposed model, the coupling between elastic deformation and photo-thermal responses was neglected. This coupling can also be included for more accuracy. When light is absorbed, a more sophisticated model can further take into account the ratio between photoactive and non-photoactive mesogens. Other geometries can be considered as well. Extensive experimental testing should be performed to help establish the best modelling approach.

%%%%%%%%%%%%%%%%%%

\begin{acks}
	The author thanks Professor Peter Palffy-Muhoray (Advanced Materials and Liquid Crystal Institute, Kent State University, Ohio, USA) for useful and stimulating discussions.
\end{acks}

%%%%%%%%%%%%%%%%%%%%%%%%%%%%%%%%%%%%%%%%%%%%%%%%%%%%%%%%%%%%
%%%%%%%%%%%%%%%%%%%%   NEW SECTION  %%%%%%%%%%%%%%%%%%%%%%%%
%%%%%%%%%%%%%%%%%%%%%%%%%%%%%%%%%%%%%%%%%%%%%%%%%%%%%%%%%%%%
%\appendix
\section{Appendix A. Shear stripes formation when pre-stretching}\label{NLC:sec:append:sse}
\setcounter{equation}{0}
\renewcommand{\theequation}{A.\arabic{equation}}
\setcounter{figure}{0}
\renewcommand{\thefigure}{A.\arabic{figure}}

In this appendix, we determine the stretch interval where shear stripes can form in the nematic LCE described by the elastic strain-energy function given by equation \eqref{NLC:eq:Wel:cp:nh}. Setting the nematic director in the relaxed and stretched configuration, respectively, as follows,
\begin{equation}\label{NLC:eq:n0ntheta}
\textbf{n}_{0}=
\left[
\begin{array}{c}
1\\
0\\
0
\end{array}
\right],\qquad
\textbf{n}=
\left[
\begin{array}{c}
\cos\theta\\
\sin\theta\\
0
\end{array}
\right],
\end{equation}
where $\theta\in[0,\pi/2]$ is the angle between $\textbf{n}$ and $\textbf{n}_{0}$, the associated natural deformation tensors, given by equation \eqref{NLC:eq:G} are, respectively,
\begin{equation}
\textbf{G}_{0}=
\left[
\begin{array}{ccc}
a^{1/3} &  0 & 0 \\
0 & a^{-1/6} & 0 \\
0 & 0 & a^{-1/6}
\end{array}
\right]
\end{equation}
and
\begin{equation}
\textbf{G}=
\left[
\begin{array}{ccc}
a^{-1/6} +\left(a^{1/3}-a^{-1/6}\right)\cos^{2}\theta &  \left(a^{1/3}-a^{-1/6}\right)\sin\theta\cos\theta & 0 \\
\left(a^{1/3}-a^{-1/6}\right)\sin\theta\cos\theta  & a^{-1/6} +\left(a^{1/3}-a^{-1/6}\right)\sin^{2}\theta & 0 \\
0 & 0 & a^{-1/6}
\end{array}
\right].
\end{equation}
To demonstrate shear-striping instability, we consider the following perturbed deformation gradient 
\begin{equation}
\textbf{F}=\left[
\begin{array}{ccc}
\lambda^{-1/2} &  0 & 0 \\
\varepsilon & \lambda & 0 \\
0 & 0 & \lambda^{-1/2}
\end{array}
\right],
\end{equation}
where $\lambda>1$ is the stretch ratio in the direction of the applied tensile force, and $0<\varepsilon\ll 1$ is a small shear parameter. The elastic deformation tensor $\textbf{A}=\textbf{G}^{-1}\textbf{F}\textbf{G}_{0}$ is then equal to
\begin{equation}
\begin{split}
\textbf{A}&=\left[
\begin{array}{ccc}
\lambda^{-1/2}\left(a^{1/2}\sin^{2}\theta+\cos^{2}\theta\right)  & \lambda\left(a^{-1/2}-1\right)\sin\theta\cos\theta & 0 \\
\lambda^{-1/2}\left(1-a^{1/2}\right)\sin\theta\cos\theta & \lambda\left(a^{-1/2}\sin^{2}\theta+\cos^{2}\theta\right) & 0 \\
0 & 0 & \lambda^{-1/2}
\end{array}
\right]\\
&+\varepsilon\left[
\begin{array}{ccc}
\left(1-a^{1/2}\right)\sin\theta\cos\theta & 0 & 0 \\
\left(\sin^{2}\theta+a^{1/2}\cos^{2}\theta\right) & 0 & 0\\
0 & 0 &0
\end{array}
\right].
\end{split}
\end{equation}
The eigenvalues $\{\lambda_{1}^2,\lambda_{2}^2,\lambda_{3}^2\}$ of the Cauchy-Green tensor $\textbf{F}\textbf{F}^{T}$ and $\{\alpha_{1}^2,\alpha_{2}^2,\alpha_{3}^2\}$ of the tensor $\textbf{A}\textbf{A}^{T}$ satisfy the following relations, respectively,
\begin{equation}
\lambda_{1}^2+\lambda_{2}^2+\lambda_{3}^2=\lambda^2+2\lambda^{-1}+\varepsilon^2
\end{equation}
and
\begin{equation}
\begin{split}
\alpha_{1}^{2}+\alpha_{2}^{2}+\alpha_{3}^{2}&=\left[\lambda^{-1/2}\left(a^{1/2}\sin^{2}\theta+\cos^{2}\theta\right)+\varepsilon\left(1-a^{1/2}\right)\sin\theta\cos\theta\right]^2\\ 
&+\left[\lambda^{-1/2}\left(1-a^{1/2}\right)\sin\theta\cos\theta+\varepsilon\left(\sin^{2}\theta+a^{1/2}\cos^{2}\theta\right)\right]^2\\
&+\left[\lambda\left(a^{-1/2}-1\right)\sin\theta\cos\theta\right]^2+\left[\lambda\left(a^{-1/2}\sin^{2}\theta+\cos^{2}\theta\right) \right]^2+\lambda^{-1}.
\end{split}
\end{equation}

%%%%%%%%%%%%%%%
\begin{figure}[htbp] 
	\begin{center}
		(a) \includegraphics[width=0.45\textwidth]{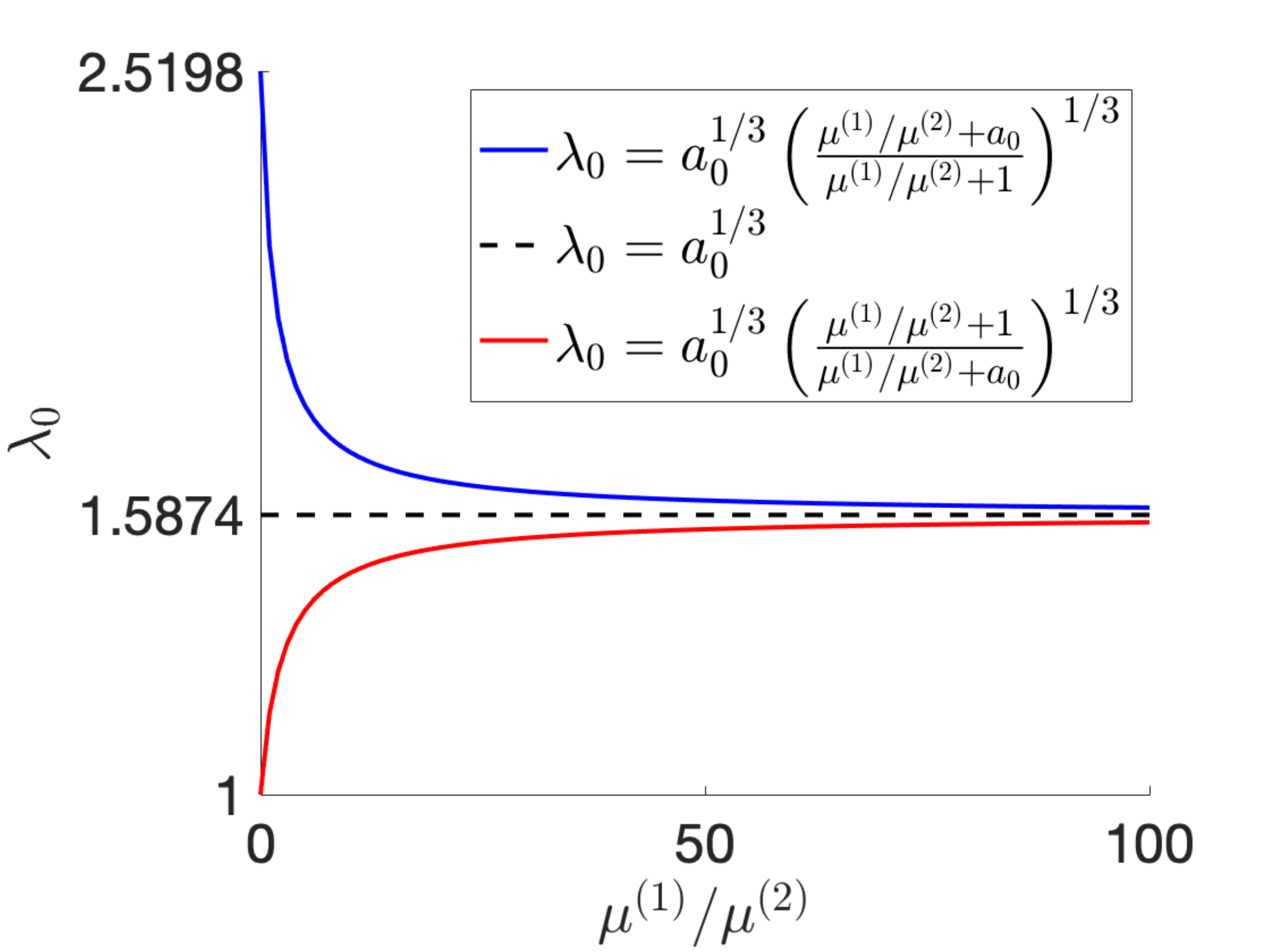}
		(b) \includegraphics[width=0.45\textwidth]{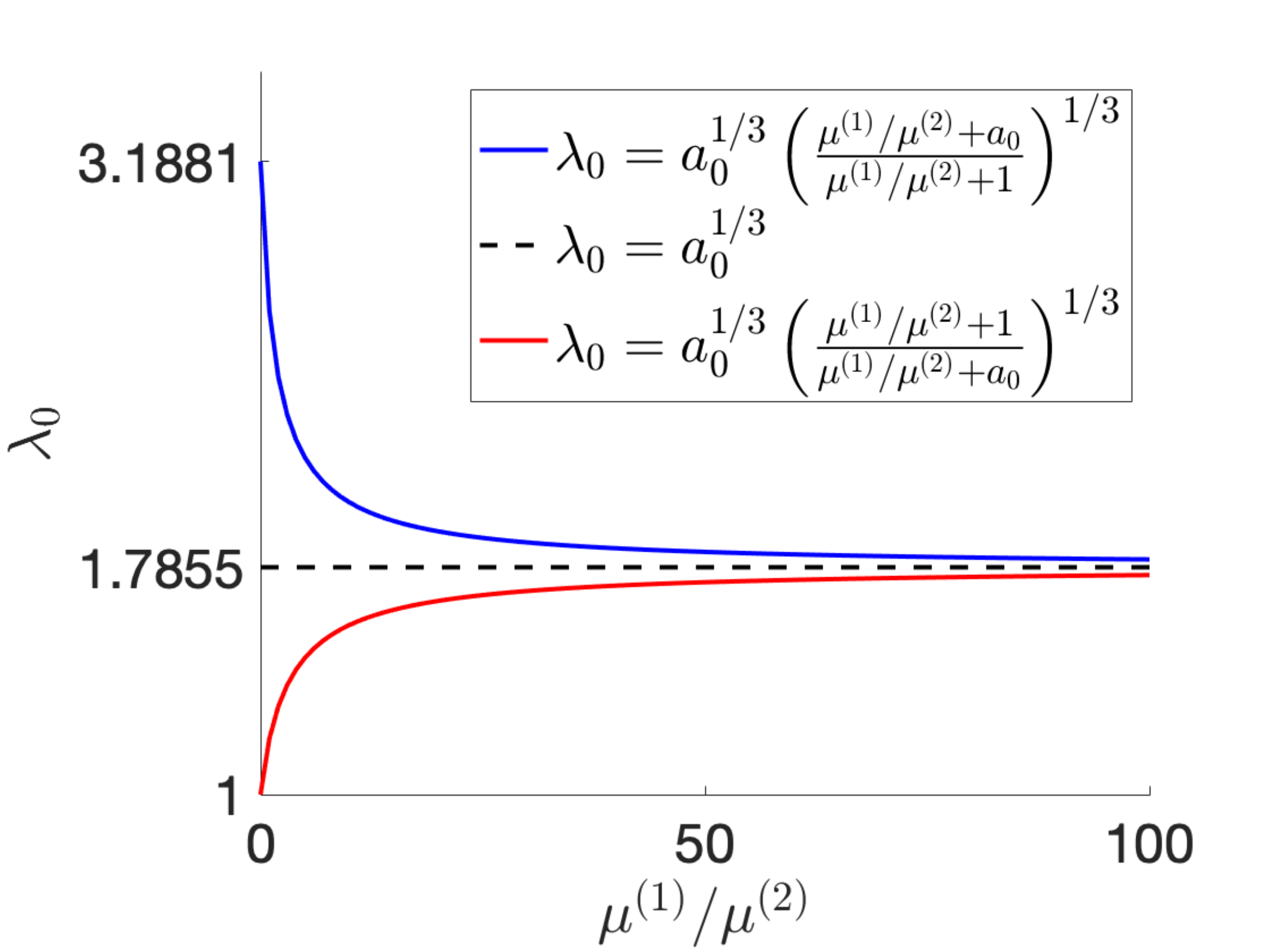}
		\caption{The stretch ratio interval where shear striping occurs in the LCE dielectric pre-stretched with ratio $\lambda_{0}$, perpendicular to the director, while the uniaxial order parameter is equal to: (a) $Q_{0}=0.5$ or (b) $Q_{0}=0.61$.}\label{NLC:fig:bounds:cp}
	\end{center}
\end{figure}
%%%%%%%%%%%%%%%

We define the following function,
\begin{equation}\label{NLC:w:epstheta}
w(\lambda,\varepsilon,\theta)=\mathcal{W}^{(el)}(\lambda_{1},\lambda_{2},\lambda_{3},\theta,a),
\end{equation}
with $\mathcal{W}^{(el)}(\lambda_{1},\lambda_{2},\lambda_{3},\theta,a)=W^{(el)}$ described by equation \eqref{NLC:eq:Wel:cp:nh}. Differentiating the above function with respect to $\varepsilon$ and $\theta$, respectively, gives
\begin{equation}\label{NLC:eq:dW:deps}
\begin{split}
\frac{\partial w(\lambda,\varepsilon,\theta)}{\partial\varepsilon}
&=\mu^{(1)}\varepsilon\\
&+\mu^{(2)}\left\{\left[\lambda^{-1/2}\left(a^{1/2}\sin^{2}\theta+\cos^{2}\theta\right)+\varepsilon\left(1-a^{1/2}\right)\sin\theta\cos\theta\right]\left(1-a^{1/2}\right)\sin\theta\cos\theta\right.\\
&\left.+\left[\lambda^{-1/2}\left(1-a^{1/2}\right)\sin\theta\cos\theta+\varepsilon\left(\sin^{2}\theta+a^{1/2}\cos^{2}\theta\right) \right]\left(\sin^{2}\theta+a^{1/2}\cos^{2}\theta\right)\right\}
\end{split}
\end{equation}
and
\begin{equation}\label{NLC:eq:dW:dtheta}
\begin{split}
\frac{\partial w(\lambda,\varepsilon,\theta)}{\partial\theta}&=\mu^{(2)}\left\{\left(a^{1/2}-1\right)\left[2\lambda^{-1/2}\sin\theta\cos\theta+\varepsilon\left(\sin^2\theta-\cos^{2}\theta\right)\right]\right.\\
&\left.\cdot\left[\lambda^{-1/2}\left(a^{1/2}\sin^{2}\theta+\cos^{2}\theta\right)+\varepsilon\left(1-a^{1/2}\right)\sin\theta\cos\theta\right]\right.\\
&\left.+\left(a^{1/2}-1\right)\left[\lambda^{-1/2}\left(\sin^{2}\theta-\cos^{2}\theta\right)-2\varepsilon\sin\theta\cos\theta\right]\right.\\
&\left.\cdot\left[\lambda^{-1/2}\left(1-a^{1/2}\right)\sin\theta\cos\theta+\varepsilon\left(\sin^{2}\theta+a^{1/2}\cos^{2}\theta\right)\right]\right.\\
&\left.+\lambda^2\left(a^{-1}-1\right) \sin\theta\cos\theta\right\}.
\end{split}
\end{equation}
The equilibrium solution minimises the energy, and thus satisfies the simultaneous equations
\begin{equation}\label{NLC:eq:dwdepstheta}
\frac{\partial w(\lambda,\varepsilon,\theta)}{\partial\varepsilon}=0\qquad\mbox{and}\qquad
\frac{\partial w(\lambda,\varepsilon,\theta)}{\partial\theta}=0.
\end{equation}
At $\varepsilon=0$ and $\theta=0$, the partial derivatives defined by equations \eqref{NLC:eq:dW:deps}-\eqref{NLC:eq:dW:dtheta} are both equal to zero. Hence, this trivial solution is always an equilibrium state. For sufficiently small values of $\varepsilon$ and $\theta$, we can write the second order approximation 
\begin{equation}\label{NLC:eq:W:2order}
w(\lambda,\varepsilon,\theta)\approx w(\lambda,0,0)+\frac{1}{2}\left(\varepsilon^2\frac{\partial^2 w}{\partial\varepsilon^2}(\lambda,0,0)+2\varepsilon\theta\frac{\partial^2 w}{\partial\varepsilon\partial\theta}(\lambda,0,0)+\theta^2\frac{\partial^2 w}{\partial\theta^2}(\lambda,0,0)\right),
\end{equation}
where
\begin{eqnarray}
&&\frac{\partial^2 w}{\partial\varepsilon^2}(\lambda,0,0)=\mu^{(1)}+\mu^{(2)}a,\\
&&\frac{\partial^2 w}{\partial\varepsilon\partial\theta}(\lambda,0,0)=\mu^{(2)}\lambda^{-1/2}\left(1-a\right),\\
&&\frac{\partial^2 w}{\partial\theta^2}(\lambda,0,0)=\mu^{(2)}\left(\lambda^2-\lambda^{-1}a\right)\left(a^{-1}-1\right).
\end{eqnarray}
First, we find the equilibrium value $\theta_{0}$ for $\theta$ as a function of $\varepsilon$ by solving the second equation in \eqref{NLC:eq:dwdepstheta}. By the approximation \eqref{NLC:eq:W:2order}, the respective equation takes the form
\begin{equation}\label{NLC:eq:thetaeps}
\varepsilon\frac{\partial^2 w}{\partial\varepsilon\partial\theta}(\lambda,0,0)+\theta\frac{\partial^2 w}{\partial\theta^2}(\lambda,0,0)=0,
\end{equation}
and implies
\begin{equation}\label{NLC:eq:theta0eps}
\theta_{0}(\varepsilon)=-\varepsilon\frac{\partial^2 w}{\partial\varepsilon\partial\theta}(\lambda,0,0)/\frac{\partial^2 w}{\partial\theta^2}(\lambda,0,0).
\end{equation}
Next, substituting $\theta=\theta_{0}(\varepsilon)$ in \eqref{NLC:eq:W:2order} gives the following function of $\varepsilon$,
\begin{equation}\label{NLC:eq:stability}
w(\lambda,\varepsilon,\theta_{0}(\varepsilon))-w(\lambda,0,0)\approx\frac{\varepsilon^2}{2}\left[\frac{\partial^2 w}{\partial\varepsilon^2}(\lambda,0,0)-\left(\frac{\partial^2 w}{\partial\varepsilon\partial\theta}(\lambda,0,0)\right)^2/\frac{\partial^2 w}{\partial\theta^2}(\lambda,0,0)\right].
\end{equation}
Depending on whether the expression on the right-hand side in \eqref{NLC:eq:stability} is positive, zero, or negative, the respective equilibrium state is stable, neutrally stable, or unstable \cite{Mihai:2020a:MG,Mihai:2021a:MG,Mihai:2023:MRGMG}. We deduce that the equilibrium state with $\varepsilon=0$ and $\theta=0$ is unstable if 
\begin{equation}\label{NLC:eq:lambda:bound1:cp}
a^{1/3}\left(\frac{\mu^{(1)}/\mu^{(2)}+1}{\mu^{(1)}/\mu^{(2)}+a}\right)^{1/3}<\lambda<a^{1/3}.
\end{equation}
Similarly, at $\varepsilon=0$ and $\theta=\pi/2$, both the partial derivatives defined by \eqref{NLC:eq:dW:deps}-\eqref{NLC:eq:dW:dtheta} are equal to zero, and
\begin{eqnarray}
&&\frac{\partial^2 w}{\partial\varepsilon^2}(\lambda,0,\pi/2)=\mu^{(1)}+\mu^{(2)},\\
&&\frac{\partial^2 w}{\partial\varepsilon\partial\theta}(\lambda,0,\pi/2)=\mu^{(2)}\lambda^{-1/2}\left(a-1\right),\\
&&\frac{\partial^2 w}{\partial\theta^2}(\lambda,0,\pi/2)=\mu^{(2)}\left(\lambda^{2}-\lambda^{-1}a\right)\left(1-a^{-1}\right).
\end{eqnarray}
Thus the equilibrium state with $\varepsilon=0$ and $\theta=\pi/2$ is unstable if 
\begin{equation}\label{NLC:eq:lambda:bound2:cp}
a^{1/3}<\lambda<a^{1/3}\left(\frac{\mu^{(1)}/\mu^{(2)}+a}{\mu^{(1)}/\mu^{(2)}+1}\right)^{1/3}.
\end{equation}
In Figure~\ref{NLC:fig:bounds:cp}, we plot the bounds given by equations \eqref{NLC:eq:lambda:bound1:cp} and \eqref{NLC:eq:lambda:bound2:cp} for the LCE dielectric pre-stretched by ratio $\lambda=\lambda_{0}$, perpendicular to the director, when $a=a_{0}$ is given by equation \eqref {NLC:eq:aQ} with $Q=Q_{0}$ and $Q_{0}=0.5$ or $Q_{0}=0.61$. For example, if $\lambda_{0}=1.25$, then shear striping cannot occur for $\mu^{(1)}/\mu^{(2)}>1.9$ when the LCE is heated and for $\mu^{(1)}/\mu^{(2)}>1.5$ when the LCE is illuminated.

%%%%%%%%%%%%%%%%%%%%%%%%%%%%%%%%%%%%%%%%%%%%%%%%%%%%%%%%%%%%
\section{Appendix B. Step-length ratios}\label{NLC:sec:append:cp}
\setcounter{equation}{0}
\renewcommand{\theequation}{B.\arabic{equation}}

We remark here that, for $\mu^{(1)}=0$, the LCE model defined by equation \eqref{NLC:eq:Wel:cp:nh} reduces to the neoclassical model considered originally in \cite{Hiscock:2011:HWPM}. Assuming that the step lengths in equation (8) there satisfy the incompressibility constraint $l_{\parallel}\left(l_{\perp}\right)^2=1$, and similarly, $l^{0}_{\parallel}\left(l^{0}_{\perp}\right)^2=1$ for the relaxed state, the step-length ratios are, respectively:
\begin{equation}
\begin{split}
p_{\parallel}&=\frac{l^{0}_{\parallel}}{l_{\parallel}}=\frac{(1+2Q)^{1/3}(1-Q)^{2/3}}{(1+2Q_{A})^{1/3}(1-Q_{A})^{2/3}}\cdot\frac{1+2Q_{A}}{1+2Q}=\frac{(1+2Q_{A})^{2/3}(1-Q)^{2/3}}{(1-Q_{A})^{2/3}(1+2Q)^{2/3}},\\
p_{\perp}&=\frac{l^{0}_{\perp}}{l_{\perp}}=\frac{(1+2Q)^{1/3}(1-Q)^{2/3}}{(1+2Q_{A})^{1/3}(1-Q_{A})^{2/3}}\cdot\frac{1-Q_{A}}{1-Q}=\frac{(1+2Q)^{1/3}(1-Q_{A})^{1/3}}{(1-Q)^{1/3}(1+2Q_{A})^{1/3}}.
\end{split}
\end{equation}
In our notation, these ratios take the equivalent form
\begin{equation}
%\begin{split}
p_{\parallel}=\frac{a_{0}^{2/3}}{a^{2/3}}=\frac{(1+2Q_{0})^{2/3}(1-Q)^{2/3}}{(1-Q_{0})^{2/3}(1+2Q)^{2/3}},\qquad
p_{\perp}=\frac{a^{1/3}}{a_{0}^{1/3}}=\frac{(1+2Q)^{1/3}(1-Q_{0})^{1/3}}{(1-Q)^{1/3}(1+2Q_{0})^{1/3}}.
%\end{split}
\end{equation}
Therefore, in the case of natural deformation under temperature changes with in-plane nematic director, an upper efficiency bound of $2.7\%$ is found in this study when $\mu^{(1)}=0$  (see Figure~\ref{NLC:fig:eff}(a)) compared to $1\%$ reported in \cite{Hiscock:2011:HWPM} .

%%%%%%%%%%%%%%%%%%%%%%%%%%%%%

%%%%%%%%%%%%%%%%%%%
\end{document}